\RequirePackage{fix-cm}
\RequirePackage{rotating}
\documentclass[smallextended]{svjour3}       % onecolumn (second format)
\smartqed  % flush right qed marks, e.g., at end of proof
\usepackage{graphicx}
\usepackage{paralist}
\usepackage{booktabs}
\usepackage{tabularx}
\usepackage{vcell}
\usepackage{multirow}
\usepackage{colortbl}
\usepackage{xcolor}
\usepackage{graphicx}
\usepackage{enumitem}
\usepackage{amsmath}
\usepackage{mdframed}
\usepackage{hyperref}
\usepackage[english]{babel}
\usepackage[autostyle=true]{csquotes}
\usepackage{tcolorbox}
\usepackage{tikz}
\usepackage{fontawesome}
\usepackage{float}
\usepackage{array}
\usepackage{makecell}
\usepackage{subcaption}
\usepackage{rotating}
\usepackage[title]{appendix}

\newcommand{\autour}[1]{\tikz[baseline=(X.base)]\node [draw=black,fill=cyan!20,thick,rectangle,rounded corners=4pt,text depth=0pt] (X) {#1};}

\renewcommand{\figurename}{Fig.}
\renewcommand{\tablename}{Tab.}
\newcommand{\blue}[1]{\textbf{\textcolor{blue}{#1}}}
\newcommand{\orange}[1]{\textbf{\textcolor{orange}{#1}}}
\newcommand{\multicell}[2][c]{\begin{tabular}[#1]{@{}c@{}}#2\end{tabular}}

\journalname{Requirements Engineering Journal}

\begin{document}

\title{Causality in Requirements Artifacts: Prevalence, Detection, and Impact}
%\subtitle{Do you have a subtitle?\\ If so, write it here}

%\titlerunning{Short form of title}        % if too long for running head

\author{Julian Frattini \and
        Jannik Fischbach \and
        Daniel Mendez \and
        Michael Unterkalmsteiner \and
        Andreas Vogelsang \and
        Krzystof Wnuk
}

\authorrunning{Frattini et al.}

\institute{J. Frattini, K. Wnuk, M. Unterkalmsteiner \at
              Blekinge Institute of Technology, Sweden \\
              \email{firstname.lastname@bth.se}
            \and
           J. Fischbach \at
              Qualicen GmbH, Germany, and University of Cologne, Germany \\
              \email{jannik.fischbach@qualicen.de}
            \and 
            D. Mendez \at
            Blekinge Institute of Technology, Sweden, and fortiss GmbH, Germany \\
             \email{daniel.mendez@bth.se}
            \and
           A. Vogelsang \at
              University of Cologne, Germany \\
              \email{vogelsang@cs.uni-koeln.de}
}

\date{Received: date / Accepted: date}
% The correct dates will be entered by the editor

\maketitle

\begin{abstract}

\textbf{Background}: Causal relations in natural language (NL) requirements convey strong, semantic information. Automatically extracting such causal information enables multiple use cases, such as test case generation, but it also requires to reliably detect causal relations in the first place. Currently, this is still a cumbersome task as causality in NL requirements is still barely understood and, thus, barely detectable.
\textbf{Objective}: In our empirically informed research, we aim at better understanding the notion of causality and supporting the automatic extraction of causal relations in NL requirements.
\textbf{Method}: In a first case study, we investigate 14.983 sentences from 53 requirements documents to understand the extent and form in which causality occurs. Second, we present and evaluate a tool-supported approach, called CiRA, for causality detection. We conclude with a second case study where we demonstrate the applicability of our tool and investigate the impact of causality on NL requirements.
\textbf{Results}: The first case study shows that causality constitutes around 28~\% of all NL requirements sentences. We then demonstrate that our detection tool achieves a macro-F\textsubscript{1} score of 82~\% on real-world data and that it outperforms related approaches with an average gain of 11.06~\% in macro-Recall and 11.43~\% in macro-Precision. Finally, our second case study corroborates the positive correlations of causality with features of NL requirements. 
\textbf{Conclusion}: The results strengthen our confidence in the eligibility of causal relations for downstream reuse, while our tool and publicly available data constitute a first step in the ongoing endeavors of utilizing causality in RE and beyond.

\keywords{Causality \and Multi Case Study \and Requirements Engineering \and Natural Language Processing}
% \PACS{PACS code1 \and PACS code2 \and more}
% \subclass{MSC code1 \and MSC code2 \and more}
\end{abstract}

\section{Introduction}
\label{sec:intro}

\paragraph{Motivation}
Sentences containing causal relations, for example \enquote{A confirmation message shall be shown if the system has successfully processed the data.}, are often used to capture the intended behavior of a system. In fact, causal relations are inherently embedded in many textual descriptions of requirements. Both understanding the extent of use, but also detecting and reliably extracting these causal relations, provide great potential for applications in the domain of Requirements Engineering (RE). Among these are, for instance, supporting the automated test case generation~\cite{frattini2020,fischbach20} or facilitating reasoning about inter-requirements dependencies~\cite{fischbach2020}. However, the automated extraction of causal relations from requirements is still challenging for two reasons: On the one hand, even though controlled natural language in RE~\cite{fuchs1995controlled,mavin2009ears} aims to minimize ambiguity and can easily be reused for further formalization~\cite{selway2015formalising,gordon2009generating}, unrestricted natural language is still predominantly used in RE~\cite{wagner2019status}. This complicates information retrieval from requirements due to the inherent complexity and ambiguity of NL. On the other hand, causal relations can occur in different forms~\cite{blanco08} such as \textit{marked}/\textit{unmarked} (i.e., containing a cue phrase indicating the causal relationship) or \textit{explicit}/\textit{implicit} (i.e., explicitly stating both the cause and the effect), further rendering the identification and extraction of causes and effects cumbersome. Existing approaches still fail to extract causal relations from NL with a performance that allows for efficient and reliable use in practice~\cite{Asghar16}. We therefore argue that a novel method for detecting and extracting causal relations from requirements is imperative for the effective utilization of causality in RE. 

\paragraph{Contribution}
Causality extraction entails two distinct challenges: first, one needs to determine whether a requirement contains causal relations. Only sentences containing causal relations are eligible for extraction, so sentences containing no causal relations can be discarded. Second -- if they contain causal relations -- these need to be properly understood and extracted. Addressing both challenges requires comprehending to which extent, form, and complexity causality occurs in practice in RE. Reliable knowledge about the distribution of causality is a necessary precondition to develop efficient approaches for the automated detection and extraction of causal relations. However, empirical evidence on causality in requirements artifacts is still unavailable to this day. In this manuscript, we report on how we addressed those challenges to close the existing research gap by making the following contributions (C):

\begin{compactitem}
  \item[C 1] \textbf{Prevalence of Causality}: We report on an exploratory case study analyzing the extent, form, and complexity of causality in requirements rooted in 14,983~sentences and emerging from 53~requirement documents, which originate from 18~different domains. We corroborate that causality tends to occur predominantly in \textit{explicit} and \textit{marked} form, and that about 28~\% of the analyzed sentences contain causal information about the expected system behavior. This strengthens our confidence in the relevance of causality and, in consequence, of our approach to automatically extract causality.
  \item[C 2] \textbf{Automated Detection of Causality}: We present our tool-supported approach CiRA (\textbf{C}ausality \textbf{i}n \textbf{R}equirement \textbf{A}rtifacts), which constitutes a first step towards causality extraction from NL requirements by automatically detecting causal relations in NL requirements. We train and empirically evaluate CiRA using the pre-analyzed data set and achieve a macro-F\textsubscript{1} score of 82~\%. Compared to baseline systems that classify causality relying on the presence of certain cue phrases, or shallow ML models, CiRA leads to an average performance gain of 11.43~\% in macro-Precision and 11.06~\% in macro-Recall. 
  \item[C 3] \textbf{Impact of Causality}: We report on a second exploratory case study evaluating the correlation between the occurrence of causality and its effects on the requirements life cycle, not only demonstrating a possible use case of the automatic causality detection approach and tool but also corroborating the positive impact on causality on the requirements process.
  \item[C 4] \textbf{Open Data and Source Code}: To strengthen the transparency and, thus, the credibility of our research, but also to facilitate independent replications, we publicly disclose our tool, code, and data set used in the case study. A demo of CiRA can be accessed at \url{http://www.cira.bth.se/bert}. Our code and annotated data sets can be found at \url{https://doi.org/10.5281/zenodo.5596668}.
\end{compactitem}

\subsection{Previously published material}
This manuscript extends our previously published conference paper~\cite{fischbach2021automatic} in the following aspects: we extend our case study (C1) and development of our own approach (C2) by the aforementioned second case study (C3) based on an extensive data set of requirements from a multinational software development company. In addition, we expand the evaluation of the resulting data from the first case study in response to discussions with the requirements engineering community to increase the generalizability of our results. Please note that we took the liberty of intentionally reusing minor parts of our previously published material for this manuscript at hand in a verbatim manner, such as discussions on related work or terminological definitions.

\subsection{Outline}
The manuscript is structured as follows: Sect.~\ref{sec:terminology} introduces the terminology used throughout the manuscript based on established literature. Sect.~\ref{sec:study} reports on the first case study investigating the extent, form, and complexity to which causality is used in NL requirements (C1). Our approach for automatically detecting causal requirements (C2) is presented in Sect.~\ref{sec:detection}. Sect.~\ref{sec:effects} reports on the second case study (C3) investigating the impact of causality in natural language requirements on their life cycle. Sect.~\ref{sec:related}, finally, presents related work in the research area before concluding our work with Sect.~\ref{sec:conclusion}.

\section{Terminology}
\label{sec:terminology}
The concept of causality has received notable attention in the studies of various disciplines, e.g., in psychology~\cite{psycho03}. Before investigating the extent to which causality occurs in requirements, we elaborate on a definition of causality.

\paragraph{Concept of Causality} 
Causality describes a relation between at least two events: a causing event (the \emph{cause}) and a caused event (the \emph{effect}). An event is commonly defined as \enquote{any situation (including a process or state) that happens or occurs either instantaneously (punctual) or during a period of time (durative)}~\cite{mostafazadeh16}. The connection between causes and effects is counterfactual~\cite{Lewis1973a}: if a cause $c_1$ does not occur, then an effect $e_1$ cannot occur either. Consequently, a causal relation entails that the effect may only occur \emph{if and only if} the cause has occurred. If this is not the case, then the relation might be confounded and is hence not causal. This relation can be interpreted in the view of Boolean algebra as an equivalence between a cause and effect ($c_1 \Leftrightarrow e_1$): if the cause is true, the effect is true and if the cause is false, the effect is also false. The representation of a causal relationship as a logical equivalence is not entirely reflecting the nature of the relation, especially in regards to the temporal order of events, which is not determined in propositional logic. The challenges of formalizing causal relationships both regarding the used notation and the ambiguity when interpreting these relationships are discussed in depth in a different paper~\cite{fischbach2021practitioners}. For the remainder of this manuscript, we use the notation of a logical equivalence to represent a causal relationship. We refer the reader interested in an extended discussion on the logical formalization of causal relationships to our previous publication on the matter~\cite{fischbach2021practitioners}. The type of causal relation between the two events can take one of three different forms~\cite{Wolff07}: a \textit{causing}, \textit{enabling}, or \textit{preventing} relationship.
\begin{compactitem}
    \item \textbf{$c_1$ causes $e_1$}: If $c_1$ occurs, $e_1$ also occurs ($c_1 \Leftrightarrow e_1$). This can be illustrated by REQ 1: \enquote{After the user enters a wrong password, a warning window shall be shown.} In this case, the wrong input is the trigger to display the window.
    \item \textbf{$c_1$ enables $e_1$}:  If $c_1$ does not occur, $e_1$ does not occur either ($e_1$ is not enabled, ($\neg c_1 \Leftrightarrow \neg e_1$)). REQ 2: \enquote{As long as you are a student, you are allowed to use the sport facilities of the university.} Only the student status enables to do sports on campus. 
    \item \textbf{$c_1$ prevents $e_1$}: If $c_1$ occurs, $e_1$ does not occur ($c_1 \Leftrightarrow \neg e_1$). REQ 3: \enquote{Data redundancy is required to prevent a single failure from causing the loss of collected data.} There will be no data loss due to data redundancy.
\end{compactitem}

\paragraph{Temporal Ordering of Causes and Effects}
Causes and effects can be related to each other in three temporal ways~\cite{mostafazadeh16}. In the first temporal relation, the cause occurs \textit{before} the effect (\emph{before} relation). In REQ 1, the user has to enter a wrong password before the warning window will be displayed. In this example, the cause and effect represent two punctual events. In the second temporal relation, the occurrence of the cause and effect \textit{overlaps}: \enquote{The fire is burning down the house.} In this case, the occurrence of the emerging fire overlaps with the occurrence of the increasingly brittle house (\emph{overlaps} relation). In the third temporal relation (\emph{during} relation), cause and effect occur simultaneously. REQ 2 describes such a relation: the effect -- being allowed to do sports on the campus -- is only valid as long as one has the student status. The start and end time of the cause is therefore also the start and end of the effect. Here, both events are durative.

\paragraph{Forms of Causality} 
The form in which causality can be expressed has three further characteristics~\cite{blanco08}: \textit{marked} and \textit{unmarked} causality, \textit{explicit} and \textit{implicit} causality, and \textit{ambiguous} and \textit{non-ambiguous} regarding its cue phrases, a linguistic concept commonly used when dealing with causality in natural language~\cite{girju2002text,chang2004causal}. A cue phrase is defined as \enquote{a word, a phrase, or a word pattern, which connects one event to the other with some relation}~\cite{chang2004causal} and therefore a lexical indicator for the causal relationship.

\begin{compactitem}
    \item \textbf{Marked and unmarked}: A causal relation is \textit{marked} if a certain cue phrase indicates its causality. The requirement \enquote{If the user presses the button, a window appears} is \textit{marked} by the cue phrase \enquote{if}, while \enquote{The user has no admin rights. He cannot open the folder.} is \textit{unmarked}.
    \item \textbf{Explicit and implicit}: An \textit{explicit} causal relation contains information about both the cause and effect. The requirement \enquote{In case of an error, the systems prints an error message to the console} is \textit{explicit} since it contains both the cause (error) and effect (error message). \enquote{A parent process kills a child process} is \textit{implicit} because the effect that the child process is terminated is not explicitly stated. Implicitly causal sentences are particularly hard to process and might be a potential source of ambiguity in RE due to their obscuring nature.
    \item \textbf{Ambiguous and non-ambiguous cue phrases}: Due to the specificity of most cue phrases in \textit{marked} causality, it seems feasible to deduce the classification of a sentence as containing causality based on the occurrence of certain cue phrases. However, certain cue phrases (e.g., since) indicate causality, but also occur in other contexts (e.g., denoting time constraints). Such cue phrases are called \textit{ambiguous}, while cue phrases (e.g., because) that predominantly indicate causality are called \textit{non-ambiguous}.
\end{compactitem}

\paragraph{Complexity of Causality}
All previous examples use the complexity-wise simplest form of causality where the causal relation consists of one single cause and one single effect. Due to the increasing complexity of systems, however, the expected behavior is described by multiple causes and effects that are connected to each other. They can be linked either by conjunctions ($c_1 \land c_2 \land \dots \Leftrightarrow e_1$) or disjunctions ($c_1 \lor c_2 \lor \dots \Leftrightarrow e_1$) or a combination of both. Furthermore, the constituents of causal relations can be contained in more than one sentence, which is a significant challenge for causality extraction as it increases the scope of causality detection beyond one single sentence. Therefore we also consider \textit{two-sentence causality}. However, causal relations scattered over more than two adjacent sentences are not considered in this research work. Additionally, the complexity increases when several causal relations are linked together, i.e., if the effect of a relation $r_1$ represents a cause in another relation $r_2$. We define such causal relations, where $r_2$ is dependent on $r_1$, as \textit{event chains} (e.g., $r_1: c_1 \Leftrightarrow e_1$ and $r_2: e_1 \Leftrightarrow e_2$).

\section{Case Study 1: Prevalence of Causality in Requirement Artifacts}
\label{sec:study}
We designed and conducted the case study following the well-established guidelines of Runeson and Höst~\cite{Runeson09}. Our case study is of exploratory nature based on the classification of Robson~\cite{Robson2002}, as we aim to unravel new insights into causality in requirement artifacts. In this section, we describe our \textit{research questions}, \textit{study objects}, \textit{study design}, \textit{study results}, and \textit{threats to validity}. We conclude by giving an overview of the \textit{implications of the study} on causality detection and extraction from NL requirements.

\subsection{Research Questions}
\label{sec:study:rqs}
Our goal in understanding the prevalence of causality in requirements artifacts encompasses the extent, form, and complexity of causality. Based on the terminology previously established in Sect.~\ref{sec:terminology}, we investigate the following research questions (RQ) in the scope of this first case study:
\begin{compactitem}
  \item[\textbf{RQ 1}] To which degree does causality occur in requirement documents?
  \item[\textbf{RQ 2}] How often do the relations \textit{cause}, \textit{enable}, and \textit{prevent} occur?
  \item[\textbf{RQ 3}] How often do the temporal relations \textit{before}, \textit{overlap}, and \textit{during} occur?
  \item[\textbf{RQ 4}] In which form does causality occur in requirement documents?
  \begin{compactitem}
      \item RQ 4a: How often does \textit{marked} and \textit{unmarked} causality occur? 
      \item RQ 4b: How often does \textit{explicit} and \textit{implicit} causality occur?
      \item RQ 4c: Which causal cue phrases are used? Are they mainly \textit{ambiguous} or \textit{non-ambiguous}?
  \end{compactitem}
  \item[\textbf{RQ 5}] At which complexity does causality occur in RE documents?
  \begin{compactitem}
      \item RQ 5a: How often do multiple causes occur?
      \item RQ 5b: How often do multiple effects occur?
      \item RQ 5c: How often does \textit{two-sentence causality} occur?
      \item RQ 5d: How often do \textit{event chains} occur?
  \end{compactitem}
  \item[\textbf{RQ 6}] Is the distribution of labels in all categories domain-independent?
\end{compactitem}

\subsection{Study Objects}
\label{sec:study:objects}
\begin{figure*}
\centering
    \includegraphics[width=\textwidth]{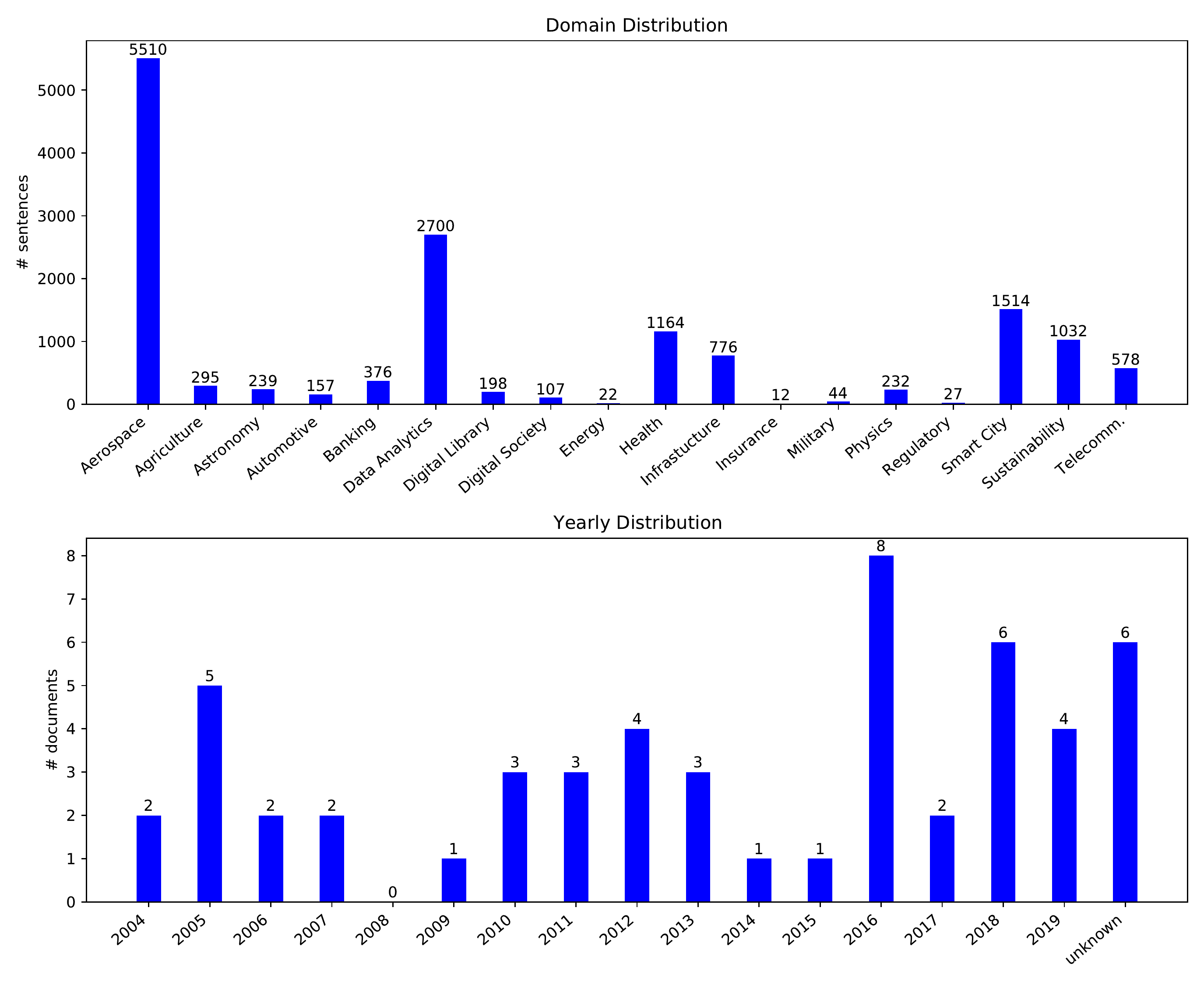}
    \vspace{-0.5cm}
    \caption{Descriptive statistics of our data set. The upper graph shows the number of sentences per domain. The lower graph depicts the year of creation per document.}
    \label{fig:dataset}
\end{figure*}

To obtain evidence on the extent to which causality is used in requirements artifacts in practice, we had to generate a large and representative collection of artifacts. We considered data sets as eligible for our case study based on three criteria: 1) the data set shall contain requirements artifacts that are/were used in practice, 2) the data set shall not be domain-specific, but rather contain artifacts from different domains, and 3) the documents shall originate from a time frame of at least 10 years. Following these criteria ensures that our analysis is not restricted to a single year or domain, but rather allows for a comprehensive and generalizable view on causality in requirements. We accordingly selected the data set provided by Fischbach et al.~\cite{fischbach2020}. To the best of our knowledge, this data set is currently the most extensive collection of requirements available to the research community. From its 463 documents containing 212k extracted and pre-processed sentences, we randomly selected 53 documents from the data set for our analysis. Our final data set consists of 14,983 sentences from 18 different domains (see Fig.~\ref{fig:dataset}). 

\subsection{Study Design}
\label{sec:study:design}

\paragraph{Model the phenomenon}
Answering the research questions in the scope of our first case study entails to annotate the sentences of our data set with respect to the categories elicited in Sect.~\ref{sec:terminology}. For example, each causal sentence has to be classified in the category \textit{Explicit} as either \textit{explicit} or \textit{implicit}. According to Pustejovsky and Stubbs~\cite{Pustejovsky12}, the first step in each annotation process is to \enquote{model the phenomenon} that needs to be annotated. Specifically, the phenomenon should be defined as a model \emph{M} that consists of a vocabulary \emph{T}, the relations \emph{R} between the terms as well as the interpretations \emph{I} of terms. RQ 1 can be understood as a binary annotation problem, which can be modeled as:
\begin{compactitem}
    \item \textbf{T}: \{sentence, causal, not causal\}
    \item \textbf{R}: \{sentence ::= causal $\mid$ not causal\}
    \item \textbf{I}: \{causal = “A sentence is causal if it contains a relation between at least two events, where e1 causes the occurrence of e2”, $\neg\text{causal}$ = “A sentence is not causal if it describes a state that is independent on any events"\}
\end{compactitem}
Explicitly modeling an annotation problem according to the aforementioned framework does not only contribute an unambiguous definition of the research problem but can also be used as a guideline for the annotators explaining the meaning of the labels. Each RQ has been modeled accordingly and communicated with all annotators. In addition to interpretation \emph{I}, we have also provided an example for each label to avoid misunderstandings. The following nine categories emerged in the process of modeling RQ 1-5, according to which we annotated our data set: \autour{Causality}, \autour{Explicit}, \autour{Marked}, \autour{Single Sentence}, \autour{Single Cause}, \autour{Single Effect}, \autour{Event Chain}, \autour{Relationship} and \autour{Temporality}. We refer to all categories except \autour{Causality} as \textit{dependent} categories, as they are dependent on the \autour{Causality} label. To answer RQ 6, we perform a stratified analysis for each of the aforementioned categories using the domains as strata. Due to the imbalance of the data set in respect to the domains the requirements sentences originate from, we formulate the following null hypothesis for each category X: \textit{sentences from different domains have the same distribution of values in category X}. 

\paragraph{Annotation Environment}
We developed our own annotation platform tailored to our research questions.\footnote[2]{A read-only view of the platform can be accessed at \url{clabel.diptsrv003.bth.se}.} In contrast to other annotation platforms~\cite{neves19} which only show single sentences to the annotators, we also display the predecessor and successor of each sentence, which is required to determine whether the causal relation is not confined to one sentence, but extends across two (see RQ 5c). For the binary annotation problems (see RQ 1, RQ 4a, RQ 4b, RQ 5a - d), we provide two labels for each category. Cue phrases present in the sentence can be manually selected by either choosing from a list of already identified cue phrases or by adding a new cue phrase using a text input field (see RQ 4c). Since RQ 2 and RQ 3 are ternary annotation problems, the platform provides three labels for these categories. 

\paragraph{Annotation Guideline} 
To ensure a common understanding both of causality itself and of the respective categories, we conducted a workshop with all annotators prior to the labeling process. The results of the workshop were recorded in the form of an annotation guideline. All annotators were instructed to comply with all of the annotation rules. One important, initially counter-intuitive instruction was to not entirely depend on the occurrence of cue phrases, as this approach is prone to introducing too many False Positives. Rather than focusing on lexical or syntactic attributes, the annotation process has to be initiated by fully reading the sentence and comprehending it on a semantic level. The impact of this becomes evident when considering some examples: requirements like \enquote{If the gaseous nitrogen supply is connected to the ECS duct system, ECS shall include the capability of monitoring the oxygen content in the ducting.} are easy to classify as causal due to the occurrence of the cue phrase \textit{if} and the explicit phrasing of both the cause and the effect. Requirements containing a relative clause like \enquote{Any items or issues which will limit the options available to the platform developers should be described.} are more difficult to correctly classify due to the lack of cue phrases. The semantically equivalent paraphrase \enquote{If an item or issue will limit the options available to the platform developers, the item or issue should be described.} reveals the causal relation contained by the requirement. A second vital instruction was to check if the cause is really necessary for the effect to occur. Only if the existence of the cause is mandatory for the effect to happen, the relation can be deemed causal. 

\begin{table}
    \centering
    \setlength{\extrarowheight}{0pt}
    \addtolength{\extrarowheight}{\aboverulesep}
    \addtolength{\extrarowheight}{\belowrulesep}
    \setlength{\aboverulesep}{0pt}
    \setlength{\belowrulesep}{0pt}
    \caption{Inter-annotator agreement statistics per category. The two categories \enquote{Relationship} and \enquote{Temporality} were jointly labeled by the first and second author and therefore do not require a reliability assessment.}
    \label{Tab:measures}
    \resizebox{\textwidth}{!}{\begin{tabular}{llcc|cc|cc|cc|cc|cc|cc|c} 
        \toprule
        &                  & \multicolumn{2}{c|}{\thead{Causality} }                                             & \multicolumn{2}{c|}{\thead{Explicit } }                                         & \multicolumn{2}{c|}{\thead{Marked } }                                           & \multicolumn{2}{c|}{\thead{Single \\ Sentence} }                                   & \multicolumn{2}{c|}{\thead{Single \\ Cause } }                                     & \multicolumn{2}{c|}{\thead{Single \\ Effect } }                                    & \multicolumn{2}{c|}{\thead{Event \\ Chain } }                                      & \textbf{avg.}                         \\
            \vcell{}               & \vcell{}         & \vcell{\textbf{0} }                      & \vcell{\textbf{1} }                     & \vcell{\textbf{0} }                    & \vcell{\textbf{1} }                     & \vcell{\textbf{0} }                    & \vcell{\textbf{1} }                     & \vcell{\textbf{0} }                    & \vcell{\textbf{1} }                     & \vcell{\textbf{0} }                    & \vcell{\textbf{1} }                     & \vcell{\textbf{0} }                    & \vcell{\textbf{1} }                     & \vcell{\textbf{0} }                     & \vcell{\textbf{1} }                    & \multicolumn{1}{l}{\vcell{}}          \\[-\rowheight]
        \printcellbottom       & \printcellmiddle & \printcellmiddle                         & \printcellmiddle                        & \printcellmiddle                       & \printcellmiddle                        & \printcellmiddle                       & \printcellmiddle                        & \printcellmiddle                       & \printcellmiddle                        & \printcellmiddle                       & \printcellmiddle                        & \printcellmiddle                       & \printcellmiddle                        & \printcellmiddle                        & \printcellmiddle                       & \multicolumn{1}{l}{\printcellmiddle}  \\
        \textbf{Confusion}     & \textbf{0}       & {\cellcolor[rgb]{0.753,0.753,0.753}}2034 & {\cellcolor[rgb]{0.753,0.753,0.753}}193 & {\cellcolor[rgb]{0.753,0.753,0.753}}24 & {\cellcolor[rgb]{0.753,0.753,0.753}}25  & {\cellcolor[rgb]{0.753,0.753,0.753}}1  & {\cellcolor[rgb]{0.753,0.753,0.753}}22  & {\cellcolor[rgb]{0.753,0.753,0.753}}12 & {\cellcolor[rgb]{0.753,0.753,0.753}}8   & {\cellcolor[rgb]{0.753,0.753,0.753}}41 & {\cellcolor[rgb]{0.753,0.753,0.753}}77  & {\cellcolor[rgb]{0.753,0.753,0.753}}63 & {\cellcolor[rgb]{0.753,0.753,0.753}}72  & {\cellcolor[rgb]{0.753,0.753,0.753}}450 & {\cellcolor[rgb]{0.753,0.753,0.753}}27 &                                       \\
        \textbf{Matrix}        & \textbf{1}       & {\cellcolor[rgb]{0.753,0.753,0.753}}274  & {\cellcolor[rgb]{0.753,0.753,0.753}}499 & {\cellcolor[rgb]{0.753,0.753,0.753}}39 & {\cellcolor[rgb]{0.753,0.753,0.753}}411 & {\cellcolor[rgb]{0.753,0.753,0.753}}12 & {\cellcolor[rgb]{0.753,0.753,0.753}}464 & {\cellcolor[rgb]{0.753,0.753,0.753}}17 & {\cellcolor[rgb]{0.753,0.753,0.753}}462 & {\cellcolor[rgb]{0.753,0.753,0.753}}43 & {\cellcolor[rgb]{0.753,0.753,0.753}}338 & {\cellcolor[rgb]{0.753,0.753,0.753}}46 & {\cellcolor[rgb]{0.753,0.753,0.753}}318 & {\cellcolor[rgb]{0.753,0.753,0.753}}13  & {\cellcolor[rgb]{0.753,0.753,0.753}}9  & \multicolumn{1}{l}{}                  \\
        \textbf{Agreement}     &                  & \multicolumn{2}{c|}{84.4~\%}                                                        & \multicolumn{2}{c|}{87.2~\%}                                                      & \multicolumn{2}{c|}{93.1~\%}                                                      & \multicolumn{2}{c|}{95.0~\%}                                                      &     \multicolumn{2}{c|}{76.0~\%}                                                      & \multicolumn{2}{c|}{76.4~\%}                                                      & \multicolumn{2}{c|}{92.0~\%}                                                      & 86.3~\%                                \\
        \textbf{Cohen's Kappa} &                  & \multicolumn{2}{c|}{0.579}                                                         & \multicolumn{2}{c|}{0.358}                                                       & \multicolumn{2}{c|}{0.023}                                                       & \multicolumn{2}{c|}{0.464}                                                       & \multicolumn{2}{c|}{0.261}                                                       & \multicolumn{2}{c|}{0.362}                                                       & \multicolumn{2}{c|}{0.27}                                                        & 0.331                                 \\
        \textbf{Gwet's AC1}    &                  & \multicolumn{2}{c|}{0.753}                                                         & \multicolumn{2}{c|}{0.84}                                                        & \multicolumn{2}{c|}{0.926}                                                       & \multicolumn{2}{c|}{0.945}                                                       &         \multicolumn{2}{c|}{0.645}                                                       & \multicolumn{2}{c|}{0.625}                                                       & \multicolumn{2}{c|}{0.91}                                                        & 0.806                                 \\
    \bottomrule
    \end{tabular}}
\end{table}

\paragraph{Annotation Validity}
We utilize the calculation of the inter-annotator agreement to verify the reliability of our annotations. Each of the six annotators was assigned 3,000 sentences, of which 2,500 were unique and 500 overlapping. Consequently, among the approximately 15,000 annotated sentences 3,000 were labeled by two annotators. To maximize the meaningfulness of the inter-annotator agreement, the 500 overlapping sentences were selected in batches of 100, such that every annotator had an overlap with every other annotator. Based on the overlapping sentences, we calculated the Cohen's Kappa~\cite{cohen60} measure to evaluate how well the annotators made the same annotation decision for a given category. We chose Cohen's Kappa since it is widely used for assessing inter-rater reliability~\cite{Viera05}. However, a number of statistical problems are known to exist with this measure~\cite{McHugh12}. In case of a high imbalance of ratings Cohen's Kappa is low and indicates poor inter-rater reliability even if there is a high agreement between the raters (Kappa paradox~\cite{FEINSTEIN1990}). Thus, the calculation of Cohen's Kappa is not meaningful in such scenarios. Consequently, studies~\cite{Wongpakaran13} suggest that Cohen's Kappa should always be reported together with the percentage of agreement and other paradox resistant measures (e.g., Gwet's AC1 measure~\cite{gwet}) in order to make a valid statement about the inter-rater reliability. We involved six annotators in the creation of the corpus and assessed the inter-rater reliability on the basis of 3,000 overlapping sentences, which represent about 20~\% of the total data set. We calculated all measures using the cloud-based version of AgreeStat~\cite{AgreeStat}. Cohen's Kappa and Gwet's AC1 can both be interpreted using the taxonomy developed by Landis and Koch~\cite{landis77}: values $\leq$ 0 as indicating no agreement and 0.01–0.20 as none to slight, 0.21–0.40 as fair, 0.41–0.60 as moderate, 0.61–0.80 as substantial, and 0.81–1.00 as almost perfect agreement. Table~\ref{Tab:measures} provides an overview of the confusion matrices and calculated agreement measures per category. The inter-rater agreement for the category \autour{Causality} was calculated on the basis of all 3000 overlapping sentences. Since the other categories represent specific forms of causality, we computed their inter-rater agreement only on the sentences marked as causal. Our analysis demonstrates that the inter-rater agreement of our annotation process is reliable. Across all categories, an average percentage of agreement of 86~\% was achieved. Except for the categories \autour{Single Cause} and \autour{Single Effect}, all categories show a percentage of agreement of at least 84~\%. We hypothesize that the slightly lower value of 76~\% for these two categories is caused by the fact that in some cases the annotators interpret the causes and effects with different granularity (e.g., annotators might break some causes and effects down into several sub causes and effects, while some do not). Hence, the annotations  differ slightly. The Kappa paradox is particularly evident for the categories \autour{Marked} and \autour{Event Chain}. Despite a high agreement of over 90~\%, Cohen's Kappa yields a very low value, which \enquote{paradoxically} suggests almost no or only fair agreement. A more meaningful assessment is provided by Gwet's AC1 as it did not fail in the case of prevalence and remains close to the percentage of agreement. Across all categories, the mean value is above 0.8, which indicates a nearly perfect agreement. Therefore, we assess our labeled data set as reliable and suitable for further analysis and the implementation of our causality detection approach.

\begin{figure*}
    \centering
    \includegraphics[width=\textwidth]{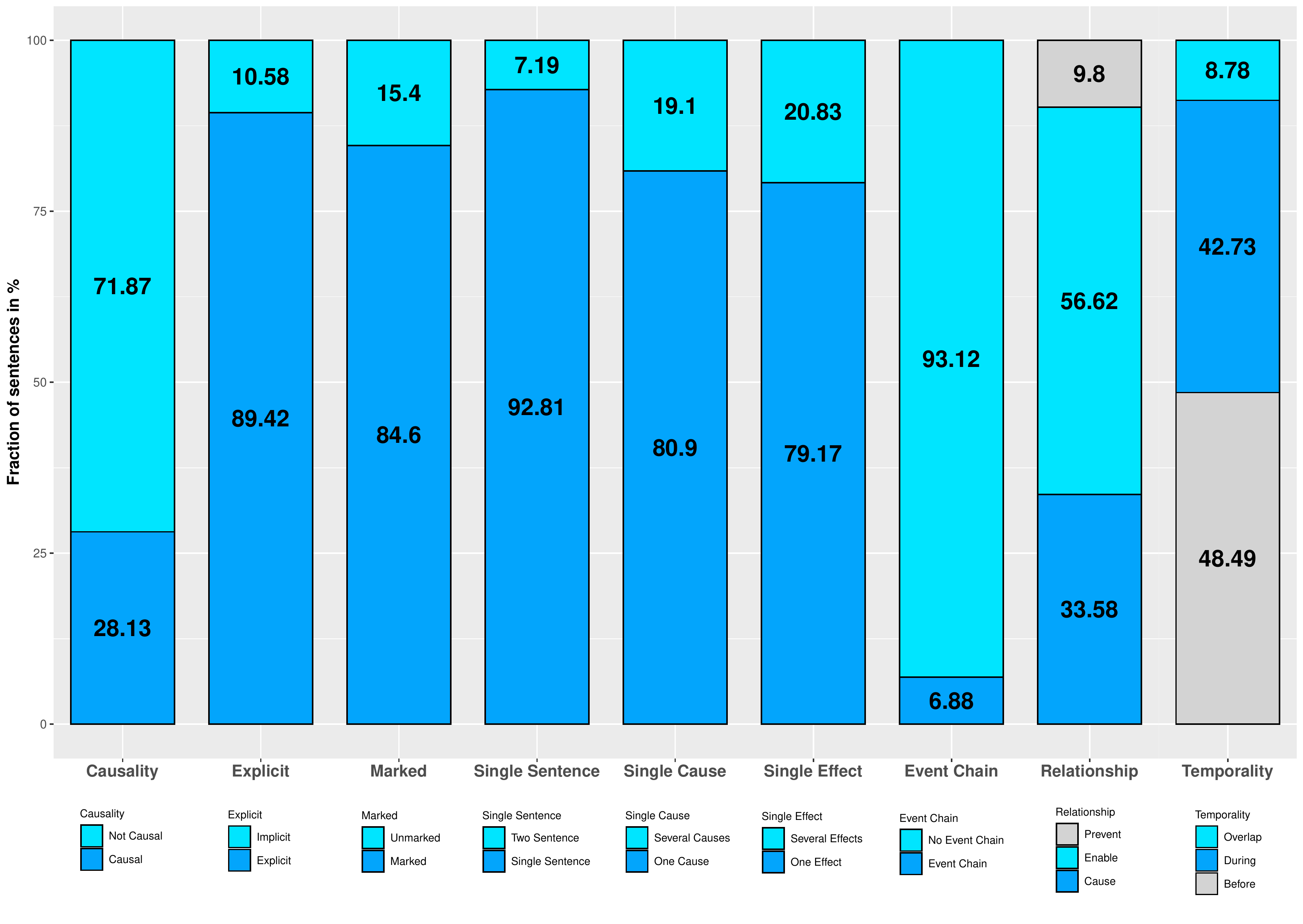}
    \vspace{-.5cm}
    \caption{Annotation results per category. The y axis of the bar plot for the category \enquote{Causality} refers to the total number of analyzed sentences. The other bar plots are only related to the causal sentences.}
    \label{fig:studyResults}
\end{figure*}

\begin{figure}
    \centering
    \includegraphics[width=0.9\textwidth]{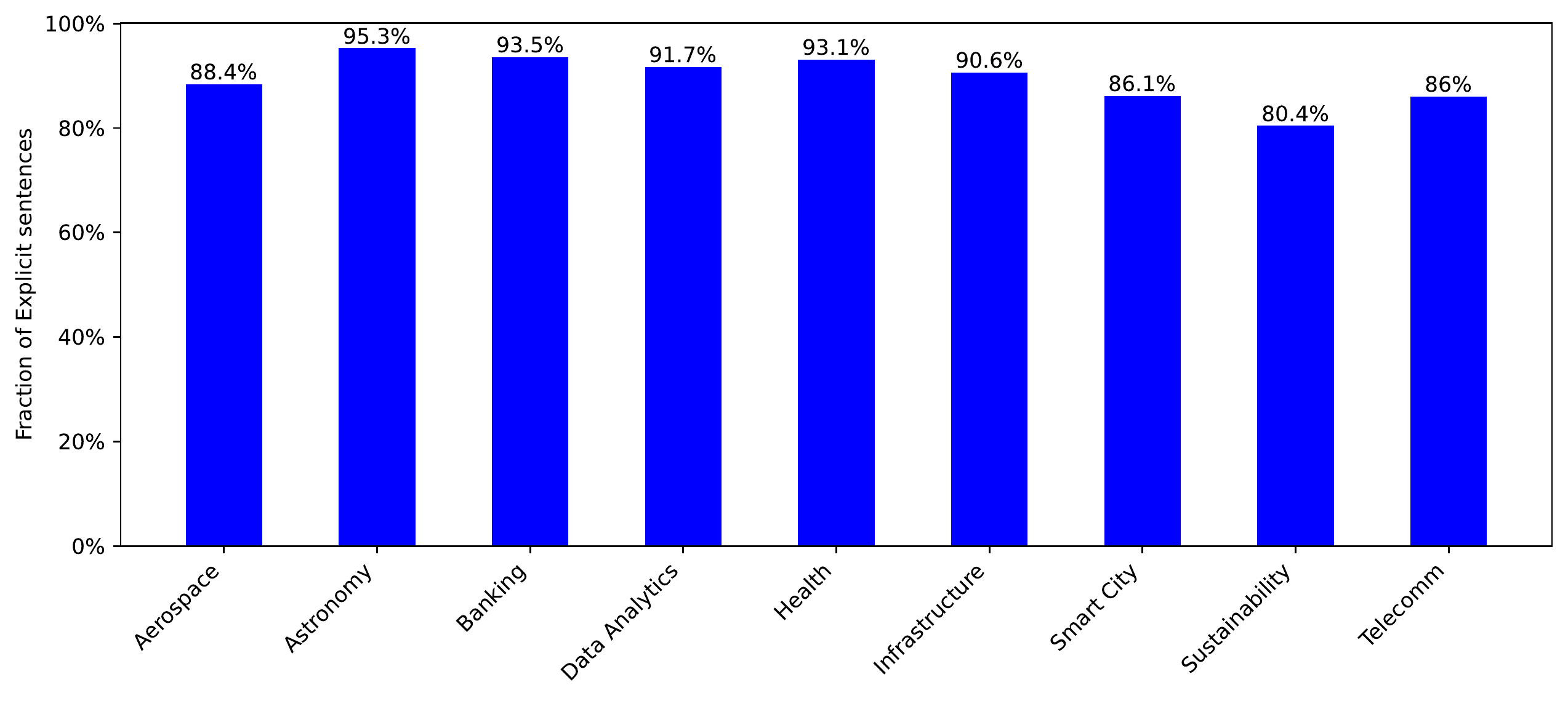}
    \vspace{-.4cm}
    \caption{Percentage of sentences labeled \enquote{Explicit} within domains containing at least 100 causal sentences}
    \label{fig:stratified:explicit}
\end{figure}

\paragraph{Data Analysis}
RQ 1-5 are answered by providing descriptive statistics of the distribution of labels for each category. For RQ 6, inferential statistics are applied. Since the hypotheses formulated for each category aim to investigate the independence between the association of a requirement to a specific domain and the distribution in the respective category, a statistical hypothesis test for independence can be used. As both the independent variable (the domain) and the dependent variable (the respective category) are categorical, the Chi-squared test will be used. The category \autour{Causality} is tested with respect to the full annotated data set. All dependent categories are tested on the causal subset of the data since only causal sentences are annotated in the other categories. For all tests, only domains with at least 100 sentences were selected as eligible strata to confine the hypothesis tests to sufficiently represented domains. This threshold was introduced to RQ 6 to avoid the noise of underrepresented domains. Since the data set was aggregated in RQ 1-5, this change is only necessary for RQ 6. Using the subset of domains as strata implies the degree of freedom of the Chi-squared tests exceeding 2, hence the risk of the multiple comparison problem, i.e., the likelihood for a Type I error in rejecting null hypotheses, arises~\cite{benjamini1995controlling}. For example, when evaluating the null-hypothesis of independence for the dichotomous category \autour{Explicit}, considering the nine eligible domains with more than 100 causal sentences yields a degree of freedom of 8, as it is calculated as follows~\cite{mchugh2013chi} (considering that the number of rows is 2 for dichotomous variables):
\begin{equation} dof = (\text{number of rows}-1) * (\text{number of columns}-1) = (2-1) * (9-1) = 8 \end{equation}
The p-value of the Chi-squared test of this hypothesis is 0.000036, far below the significance level $\alpha = 0.05$, even though the relative number of values in the category \autour{Explicit} among the eligible domains suggests an equal distribution and therefore independence of the domain, as seen in Fig.~\ref{fig:stratified:explicit}. Hence, instead of reporting the in this case not meaningful p-value of the Chi-square hypothesis test we perform a Bonferroni correction~\cite{benjamini1995controlling} on the significance level and perform the Chi-squared test in each category for each domain against the sum of all samples outside of the domain, as applied in similar scenarios~\cite{latta2012use}. Applying the Bonferroni correction to the significance level based on the following formula~\cite{benjamini1995controlling} yields a significance level that counteracts the large degree of freedom $m$ and reduces the likelihood of Type I errors when refuting null hypotheses:
\begin{equation} p_c = \frac{\alpha}{m} = \frac{0.5}{8} = 0.00625 \end{equation}
The previously calculated p-value for the Chi-square test of independence considering all domains still suggests to reject the null hypothesis. Hence, a post-hoc test similar to~\cite{latta2012use}, where each domain is compared to the sum of all other domains, is applied to reveal, that only the null-hypothesis for the domain \textit{sustainability} can be refuted with a p-value of $0.0001 < 0.00625$, which aligns with Fig.~\ref{fig:stratified:explicit}. This procedure is applied to all hypotheses of RQ 6.

\subsection{Study Results}
\label{sec:study:results}
The results of each labeled category are visualized in Fig.~\ref{fig:studyResults}. Detailed values for the distribution of labels among all categories and domains are given in Table~\ref{tab:distribution:overview}. When interpreting the values, it is important to note that for our analysis we consider complete requirement documents. Consequently, our data set contains records with different contents, which do not necessarily represent all functional requirements. For example, requirement documents also contain non-functional requirements, phrases for content structuring, purpose statements, etc. The results are hence to be interpreted in respect to the content of a full requirements artifact, not only to its functional requirements.

\textbf{Answer to RQ 1:} Fig.~\ref{fig:studyResults} confirms that causality occurs in requirement documents to a significant extent. About 28~\% of the analyzed sentences are causal. From this result, we can conclude that causality is a major linguistic element of requirement artifacts as almost one-third of all sentences are causal. 

\textbf{Answer to RQ 2:} 
The majority (56~\%) of causal sentences contained in requirement artifacts represent an \textit{enable} relationship between certain events. Only about 10~\% of the causal sentences indicate a \textit{prevent} relationship. \textit{Cause} relationships are found in about 34~\% of the annotated data. 

\textbf{Answer to RQ 3:}
Interestingly, we found that causes and effects occur almost equally often in a \textit{before} and \textit{during} relation. With about 48~\%, the \textit{before} relation is the most frequent temporal relation in our data set, but only with a difference of about 6~\% compared to the \textit{during} relation. The \textit{overlap} relation occurred only in a minority (8.78~\% of the sentences). 

\textbf{Answer to RQ 4a:} The majority of causal sentences contain one or more cue phrases, as Fig.~\ref{fig:studyResults} indicates, to indicate the causal relationship between certain events. Only around 15~\% of the labeled sentences were categorized as \textit{unmarked} causality. 

\textbf{Answer to RQ 4b:} Most causal sentences are \textit{explicit}, i.e., they contain information about both the cause and the effect. Only about 10~\% of causal sentences are \textit{implicit}.

\textbf{Answer to RQ 4c:} 
All causal cue phrases identified in the investigated requirements artifacts are listed in Tab.~\ref{Tab:cuePhrases}. The left side of the table shows the cue phrases ordered by word group. On the right side, all verbs used to express causal relations are listed. The verbs are further ordered according to whether they express a \textit{cause}, \textit{enable}, or \textit{prevent} relationship. To assess the ambiguity of a cue phrase x, we formulate a binary classification task: consider all sentences as the sample space. The causal sentences of that sample space represent the relevant elements. The precision of cue phrase x as a selection criterion for causal sentences is the conditional probability, that a sentence from the sample space is causal given that it contains cue phrase x, and hence reflects the ambiguity of the cue phrase: 
\begin{equation}
    \begin{split}
        \Pr( \text{sentence is causal} \mid \text{sentence contains x}) = \\ \frac{Pr(\text{sentence is causal} \cap \text{sentence contains x})}{Pr(\text{sentence contains x})}
    \end{split}
\end{equation}
A high precision value indicates a \textit{non-ambiguous} cue phrase, i.e., the occurrence of the cue phrase in a sentence is a strong indicator for the sentence being causal, while low values indicate strongly \textit{ambiguous} cue phrases. Tab.~\ref{Tab:cuePhrases} demonstrates that a number of different cue phrases are used to express causality in requirement documents. Not surprisingly, cue phrases like \enquote{if}, \enquote{because} and \enquote{therefore} show precision values of more than 90~\%. However, there is a variety of cue phrases that indicate causality in some sentences but also occur in other non-causal contexts. This is especially evident in the case of pronouns. Relative sentences can indicate causality, but not in every case, which is reflected by the low precision value of for example \textit{which}. A similar pattern emerges with regard to the used verbs. Only a few verbs (e.g., \enquote{leads to, degrade, and enhance}) show a high precision value. Consequently, the majority of used pronouns and verbs do not necessarily indicate a causal relation if they are present in a sentence. 

\textbf{Answer to RQ 5a:} Fig.~\ref{fig:studyResults} illustrates that most causal relations in requirement documents include only a single cause. Multiple causes occur in only 19.1~\% of the analyzed causal sentences. The exact number of causes was not documented during the annotation process. However, the participating annotators reported consistently that two to three causes were predominantly prevalent in the case of complex causal relations. More than three causes were rare. 

\textbf{Answer to RQ 5b:} Interestingly, the distribution of effects is similar to that of causes. Likewise, single effects occur significantly more often than multiple effects. According to the annotators, the number of effects in the case of complex relations is mostly limited to a maximum of two effects. Three or more effects occur rarely.

\textbf{Answer to RQ 5c:} Most causal relations can be found in single sentences. Relations, where cause and effect are distributed over two sentences, occur only in about 7~\% of the analyzed data. Among the \textit{marked} subset of these sentences ($n=242$) the cue phrase \enquote{therefore} was used most frequently, occurring 58 times. The next-most frequent cue phrase, \enquote{thus}, appeared only 14 times.

\textbf{Answer to RQ 5d:} Fig.~\ref{fig:studyResults} shows that \textit{event chains} are rarely used in requirement documents. Most causal sentences contain isolated causal relations, while only roughly 7~\% contain \textit{event chains}.

\begin{table}
    \centering
    \caption{Overview of cue phrases used to indicate causality in requirement documents. \textbf{Bold} precision values highlight non-ambiguous phrases that mostly indicated causality ($\Pr( \text{Causal} \mid \text{X is present in sentence}) \geq 0.8$).}
    \vspace{-0.2cm}
    \resizebox{\textwidth}{!}{\begin{tabular}{lllll||lllll} 
        \toprule
        \textbf{Type} & \textbf{Phrase}   & \textbf{Causal} & \textbf{Not Causal} & \begin{tabular}[c]{@{}l@{}}\textbf{Precision}\end{tabular} & \textbf{Type} & \textbf{Phrase}   & \textbf{Causal} & \textbf{Not Causal} & \textbf{Precision} \\ 
        
        \midrule
        conjunctions & if & 387 & 41  & \textbf{0.90} & Cause & force(s/ed) & 21 & 18 & 0.54 \\
        & as            & 607   & 1313  & 0.32          & & cause(s/ed)     & 32    & 10    & 0.76 \\
        & because       & 78    & 7     & \textbf{0.92} & & lead(s) to      & 5     & 0     & \textbf{1.00} \\
        & but           & 100   & 204   & 0.33          & & reduce(s/ed)    & 48    & 28    & 0.63 \\
        & in order to   & 141   & 33    & \textbf{0.81} & & minimize(s/ed)  & 28    & 11    & 0.72 \\
        & so (that)     & 88    & 86    & 0.51          & & affect(s/ed)    & 13    & 19    & 0.41 \\
        & unless        & 23    & 4     & \textbf{0.85} & & maximize(s/ed)  & 11    & 5     & 0.69 \\
        & while         & 71    & 90    & 0.44          & & eliminate(s/ed) & 8     & 11    & 0.42 \\
        & once          & 48    & 15    & 0.76          & & result(s/ed) in & 50    & 43    & 0.54 \\
        & except        & 9     & 5     & 0.64          & & increase(s/ed)  & 49    & 34    & 0.59 \\
        & as long as    & 12    & 1     & \textbf{0.92} & & decrease(s/ed)  & 5     & 8     & 0.38 \\ 
        \cline{1-5}
        adverbs & therefore & 61 & 6 & \textbf{0.91} & & impact(s) & 37 & 68 & 0.35 \\
        & when      & 331   & 64    & \textbf{0.84} & & degrade(s/ed)   & 11    & 2     & \textbf{0.85}  \\
        & whenever  & 10    & 0     & \textbf{1.00} & & introduce(s/ed) & 11    & 12    & 0.48 \\
        & hence     & 21    & 9     & 0.70          & & enforce(s/ed)   & 2     & 1     & 0.67 \\
        & where     & 213   & 150   & 0.59          & & trigger(s/ed)   & 11    & 7     & 0.61 \\ 
        & then      & 111   & 70    & 0.61          & & imply & 7 & 14 & 0.33 \\
        & since     & 65    & 32    & 0.67          & & attain(s/ed) & 3 & 13 & 0.18 \\
        & consequently      & 2     & 6     & 0.25          & & create(s/ed) & 39 & 88 & 0.30 \\
        & wherever          & 5     & 2     & 0.71          & & impose(s/ed) & 7 & 13 & 0.35 \\
        & rather            & 16    & 30    & 0.35          & & perform(s/ed) & 26 & 60 & 0.30 \\
        \cline{6-10}
        & to this/that end  & 12    & 0     & \textbf{1.00} & Enable  & depend(s) on & 28 & 21 & 0.57 \\
        & thus              & 66    & 17    & \textbf{0.80} & & require(s/ed)       & 316   & 262   & 0.55 \\
        & for this reason   & 7     & 3     & 0.70          & & allow(s/ed)         & 187   & 130   & 0.59 \\
        & due to            & 91    & 26    & 0.78          & & need(s/ed)          & 98    & 162   & 0.38 \\
        & thereby           & 4     & 2     & 0.67          & & necessitate(s/ed)   & 7     & 2     & 0.78 \\
        & as a result       & 11    & 4     & 0.73          & & facilitate(s/ed)    & 29    & 28    & 0.51 \\
        & for this purpose  & 1     & 2     & 0.33          & & enhance(s/ed)       & 16    & 4     & \textbf{0.80}  \\ 
        \cline{1-5}
        pronouns & which & 277  & 608 & 0.31                & & ensure(s/ed)        & 145   & 66    & 0.69 \\
        & who   & 28    & 52    & 0.35                      & & achieve(s/ed)       & 30    & 24    & 0.56 \\ 
        & that & 732 & 1178 & 0.38                          & & support(s/ed)       & 128   & 301   & 0.30 \\
        & whose     & 16    & 11    & 0.59                  & & enable(s/ed)        & 75    & 36    & 0.68 \\ 
        \cline{1-5}
        adjectives & only & 127 & 126 & 0.50                & & permit(s/ed) & 10 & 13 & 0.43 \\
        & prior to          & 26    & 20    & 0.57          & & rely on     & 3     & 5     & 0.38 \\
        & imperative        & 1     & 3     & 0.25          & & measure(s/ed) & 99 & 247 & 0.28 \\
        & necessary (to)    & 36    & 19    & 0.65          & & provide(s/ed) & 75 & 125 & 0.37 \\
        & given             & 73    & 140   & 0.34          & & get & 13 & 23 & 0.36 \\
        & following         & 53    & 175   & 0.23          & & meet & 42 & 34 & 0.55 \\
        \cline{1-10}
        preposition         & for   & 1209  & 2753  & 0.31  & Prevent & hinder(s/ed) & 1 & 1 & 0.50  \\
        & during            & 327   & 137   & 0.70          & & prevent(s/ed)   & 38    & 17    & 0.69 \\
        & after             & 133   & 57    & 0.70          & & avoid(s/ed) & 14 & 23 & 0.38 \\
        & by                & 506   & 1171  & 0.30          & & mitigate(s/ed) & 3 & 8 & 0.27 \\
        & with              & 680   & 1554  & 0.30          & & & & & \\
        & in the course of  & 2     & 1     & 0.67          & & & & & \\
        & through           & 114   & 204   & 0.36          & & & & & \\
        & as part of        & 19    & 51    & 0.27          & & & & & \\
        & in this case      & 18    & 3     & \textbf{0.86} & & & & & \\
        & before            & 54    & 27    & 0.67          & & & & & \\
        & until             & 33    & 11    & 0.75          & & & & & \\
        & upon              & 25    & 48    & 0.34          & & & & & \\
        & in case of        & 30    & 7     & \textbf{0.81} & & & & & \\
        & in both cases     & 1     & 0     & \textbf{1.00} & & & & & \\
        & in the event of   & 15    & 2     & \textbf{0.88} & & & & & \\
        & in response to    & 6     & 7     & 0.46          & & & & & \\
        & in the absence of & 8     & 1     & \textbf{0.89} & & & & & \\
        & within            & 150   & 315   & 0.32         & & & & & \\
        & as far as         & 4     & 5     & 0.44         & & & & & \\
        & according to      & 21    & 54    & 0.28         & & & & & \\
        & around            & 25    & 41    & 0.37         & & & & & \\
        & from              & 370   & 990   & 0.27         & & & & & \\
        & based on          & 56    & 175   & 0.24         & & & & & \\
\bottomrule
\end{tabular}}
\label{Tab:cuePhrases}
\end{table}

\begin{figure}
    \centering
    \includegraphics[width=0.9\textwidth]{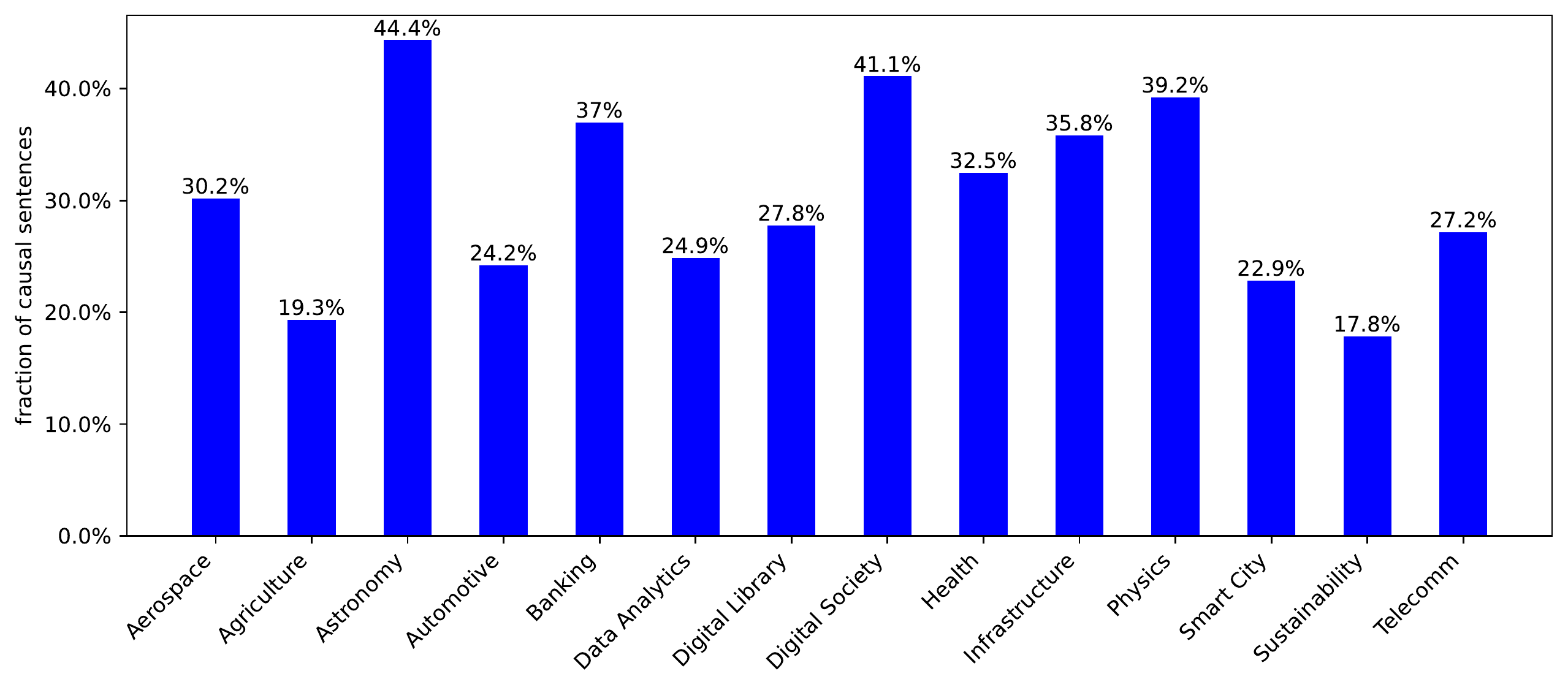}
    \vspace{-0.3cm}
    \caption{Distribution of causality among domains}
    \label{fig:domaincausality}
\end{figure}

\textbf{Answer to RQ 6} Fig.~\ref{fig:domaincausality} visualizes the distribution of causality among all domains which are represented with more than 100 sentences. As the percentage of causal sentences ranges from 17.8~\% up to 44.4~\%, we can assume that causality is indeed a phenomenon occurring in all eligible domains. The Chi-squared test reported in Tab.~\ref{tab:rq6a} suggests rejecting the null hypothesis for domain-independence for 10 out of 14 eligible domains considering the Bonferroni-corrected significance level. We can conclude that causality is a phenomenon observable independent of the domain from which requirements originate, but the extent to which causality occurs differs with statistical significance. For all dependent categories, the domains \textit{Aerospace}, \textit{Astronomy}, \textit{Banking}, \textit{Data Analytics}, \textit{Health}, \textit{Infrastructure}, \textit{Smart City}, \textit{Sustainability}, and \textit{Telecomm} are eligible for consideration as they contain more than 100 causal sentences. On the right side of Tab.~\ref{tab:rq6a} each cell contains the p-value for a Chi-squared test comparing the distribution of the given domain to the rest of the sample. Where the p-value for a given domain and category is lower than the Bonferroni-corrected significance level (denoted for each category as $p_c$), the cell is prefixed with an asterisk. The Chi-squared test of independence does not suggest to reject the null hypothesis for the categories \autour{Single Cause} and \autour{Event Chain}, but the distribution of 4 out of the eligible 9 domains in the category \autour{Temporality} are significantly different. We can conclude that the distribution of values in all categories is domain-independent to a certain degree: while the complexity of causality is mostly domain-independent, the distribution of temporality differs significantly for a about a third of the eligible domains. 
A stratified analysis for RQ 4c is reported in Tab.~\ref{tab:markers:frequent} and shows considerable differences in the usage of cue phrases in the domains, but also a degree of overlap: the cue phrase \textit{if} is among the five most frequent cue phrases in all domains, closely followed by the cue phrases \textit{when} and \textit{where}. The stratified frequencies align with the overall distribution reported in Tab.~\ref{Tab:cuePhrases} and lead to the assumption, that the distribution of cue phrases is mostly domain independent. When looking at the most precise cue phrases per domain in Tab.~\ref{tab:markers:precisemost} and the least precise cue phrases per domain in Tab.~\ref{tab:markers:preciseleast}, the cue phrases also reflect the findings from the overall distribution: precise cue phrases like \textit{if}, \textit{when}, and \textit{because} as well as infrequent, but precise causative verbs are equally represented in the domains just as imprecise cue phrases like \textit{for} or \textit{by}. We conclude that despite slight domain-specific variations, the results for RQ 4c are also domain-independent\footnote{More extensive tables reporting on the frequency and precision of cue phrases in eligible domains are included in the replication package.}.

\begin{table}
    \centering
    \caption{Bonferroni-corrected Chi-squared tests of independence from the domain. Cells prefixed with * indicate a category, where the distribution of the given domain differs significantly from the sample.}
     \vspace{-0.2cm}
    \label{tab:rq6a}
    \resizebox{\textwidth}{!}{
    \begin{tabular}{l|r|rrrrrrrr}
        \toprule
        Domain & Causal & Explicit & Marked & \multicell{Single\\Cause} & \multicell{Single\\Effect} & \multicell{Event\\Chain} & \multicell{Single\\Sentence} & Temporality & Relationship \\
        \midrule
        $p_c$             &   3.8E-03 &   6.3E-03 &   6.3E-03 &     6.3E-03 &      6.3E-03 &    6.3E-03 &                 6.3E-03 &     3.1E-03 &      3.1E-03 \\
        \midrule
        Aerospace       &  *8.0E-05 &   1.5E-01 &   8.2E-03 &     9.8E-02 &      6.4E-03 &    7.2E-02 &                 6.3E-01 &    *3.3E-10 &      3.1E-02 \\
        Agriculture     &  *7.0E-04 & \multicolumn{8}{c}{{\cellcolor[rgb]{0.753,0.753,0.753}}(domain contained less than 100 causal sentences)} \\
        Astronomy       &  *4.1E-08 &   6.2E-02 &   1.0E-02 &     8.9E-01 &      2.7E-01 &    1.1E-01 &                 2.4E-01 &    *4.4E-05 &      3.0E-01 \\
        Automotive      &   2.9E-01 & \multicolumn{8}{c}{{\cellcolor[rgb]{0.753,0.753,0.753}}(domain contained less than 100 causal sentences)} \\
        Banking         &  *1.9E-04 &   1.3E-01 &   5.3E-01 &     9.0E-02 &      2.8E-02 &    5.3E-01 &                 2.6E-01 &    *7.8E-06 &      4.4E-01 \\
        Data Analytics  &  *1.4E-05 &   3.4E-02 &   1.3E-02 &     2.3E-02 &     *3.7E-03 &    1.3E-01 &                 3.9E-01 &     3.6E-01 &      5.9E-01 \\
        Digital Library &   9.3E-01 & \multicolumn{8}{c}{{\cellcolor[rgb]{0.753,0.753,0.753}}(domain contained less than 100 causal sentences)} \\
        Digital Society &   4.4E-03 & \multicolumn{8}{c}{{\cellcolor[rgb]{0.753,0.753,0.753}}(domain contained less than 100 causal sentences)} \\
        Health          &  *9.4E-04 &   1.4E-02 &   5.7E-01 &     7.9E-02 &      6.8E-01 &    6.3E-01 &                 7.7E-01 &     7.0E-02 &      1.8E-01 \\
        Infrastructure  &  *2.1E-06 &   5.0E-01 &   3.2E-01 &     4.1E-01 &      1.6E-01 &    5.0E-01 &                 6.8E-01 &     6.1E-01 &     *6.7E-05 \\
        Physics         &  *2.1E-04 & \multicolumn{8}{c}{{\cellcolor[rgb]{0.753,0.753,0.753}}(domain contained less than 100 causal sentences)} \\
        Smart City      &  *8.4E-07 &   5.9E-02 &  *2.0E-05 &     1.3E-02 &      3.5E-01 &    3.2E-01 &                *2.3E-03 &    *3.1E-05 &      3.9E-01 \\
        Sustainability  &  *1.4E-14 &  *1.2E-04 &  *2.3E-04 &     8.7E-01 &      5.4E-01 &    1.4E-02 &                *5.7E-03 &     3.5E-01 &      1.9E-01 \\
        Telecomm        &   5.7E-01 &   2.2E-01 &   7.3E-01 &     5.2E-01 &      3.1E-01 &    1.8E-02 &                 3.5E-02 &     1.0E-01 &      7.1E-01 \\
        \bottomrule
    \end{tabular}}
\end{table}

\begin{table}
    \centering
    \caption{Distribution and precision of cue phrases in eligible domains}
    
    \begin{subtable}{1\textwidth}
    \caption{Relative frequency of cue phrases within one domain}
        \label{tab:markers:frequent}
        \resizebox{\textwidth}{!}{
        \begin{tabular}{l|rrrrr}
            \toprule
            {} &        most frequent &                 2nd most frequent &                 3rd most frequent &                 4th most frequent &                5th most frequent \\
            Domain         &                      &                     &                     &                     &               \\
            \midrule
            Aerospace      &        during (8.5\%) &         when (8.4\%) &           if (7.5\%) &  in order to (3.7\%) &       after (3.0\%) \\
            Astronomy      &          for (10.2\%) &       during (9.1\%) &        allow (9.1\%) &           in (5.7\%) &          to (5.7\%) \\
            Banking        &           if (11.2\%) &    to ensure (8.8\%) &         once (7.2\%) &        allow (5.6\%) &     through (4.8\%) \\
            Data Analytics &           if (10.7\%) &         when (9.4\%) &        where (7.9\%) &          for (5.2\%) &      during (4.2\%) \\
            Health         &         when (11.5\%) &          if (11.5\%) &      during (10.5\%) &          for (4.6\%) &       after (3.8\%) \\
            Infrastructure &        where (16.5\%) &          if (15.1\%) &    to ensure (5.6\%) &          for (5.3\%) &        then (4.9\%) \\
            Smart City     &  in order to (10.7\%) &         when (7.4\%) &           if (7.1\%) &    therefore (4.9\%) &        that (4.4\%) \\
            Sustainability &     therefore (8.6\%) &  in order to (7.0\%) &           if (5.9\%) &          for (5.4\%) &       where (4.3\%) \\
            Telecomm       &            if (8.7\%) &       during (8.0\%) &  in order to (6.0\%) &         when (5.3\%) &  in case of (4.7\%) \\
            \bottomrule
        \end{tabular}}
        \end{subtable}
        
        \bigskip
        \begin{subtable}{1\textwidth}
        \caption{Most precise cue phrases of each eligible domain}
         \label{tab:markers:precisemost}
        \resizebox{\textwidth}{!}{
        \begin{tabular}{l|rrrr}
            \toprule
            {} & most precise & 2nd most precise & 3rd most precise & 4th most precise \\
            Domain          &                             &                          &                          &                          \\
            \midrule
            Aerospace       &       imposes (100.0\%) &        as far as (100.0\%) &       result from (100.0\%) &  in this case (100.0\%) \\
            Agriculture     &         since (100.0\%) &            whose (100.0\%) &            before (100.0\%) &          when (100.0\%) \\
            Astronomy       &        during (100.0\%) &  in the event of (100.0\%) &           so that (100.0\%) &        attain (100.0\%) \\
            Automotive      &          when (100.0\%) &      in order to (100.0\%) &         therefore (100.0\%) &         whose (100.0\%) \\
            Banking         &   in order to (100.0\%) &           reduce (100.0\%) &  will be required (100.0\%) &         since (100.0\%) \\
            Data Analytics  &    as long as (100.0\%) &        increases (100.0\%) &            unless (100.0\%) &       so that (100.0\%) \\
            Digital Library &      only for (100.0\%) &            cause (100.0\%) &           because (100.0\%) &        unless (100.0\%) \\
            Digital Society &   in order to (100.0\%) &               if (100.0\%) &            due to (100.0\%) &        allows (100.0\%) \\
            Health          &        allows (100.0\%) &          so that (100.0\%) &   in the event of (100.0\%) &   as a result (100.0\%) \\
            Infrastructure  &       lead to (100.0\%) &          prevent (100.0\%) &             whose (100.0\%) &         until (100.0\%) \\
            Physics         &     is needed (100.0\%) &            after (100.0\%) &         therefore (100.0\%) &          when (100.0\%) \\
            Smart City      &     result in (100.0\%) &       to measure (100.0\%) &          increase (100.0\%) &       thereby (100.0\%) \\
            Sustainability  &  will require (100.0\%) &          enables (100.0\%) &           ensures (100.0\%) &       because (100.0\%) \\
            Telecomm        &        allows (100.0\%) &          enables (100.0\%) &            during (100.0\%) &       lead to (100.0\%) \\
                        \bottomrule
        \end{tabular}}
         \end{subtable}
        
        \bigskip
        \begin{subtable}{1\textwidth}
        \caption{Least precise cue phrases of each eligible domain}
        \label{tab:markers:preciseleast}
        \resizebox{\textwidth}{!}{
        \begin{tabular}{l|rrrr}
            \toprule
            {} &         least precise &                   2nd least precise &                     3rd least precise &                  4th least precise \\
            Domain          &                       &                     &                       &                    \\
            \midrule
            Aerospace       &  according to (14.3\%) &           to get (20.0\%) &  to mitigate (20.0\%) &  based on (24.1\%) \\
            Agriculture     &            on (26.1\%) &             that (27.1\%) &           at (28.2\%) &    allows (33.3\%) \\
            Astronomy       &          from (30.0\%) &               by (31.0\%) &       within (40.0\%) &     where (50.0\%) \\
            Automotive      &         which (10.0\%) &              for (20.0\%) &           on (24.2\%) &      that (31.8\%) \\
            Banking         &         which (21.4\%) &       to provide (25.0\%) &       create (28.6\%) &        by (28.8\%) \\
            Data Analytics  &       to meet (16.7\%) &            imply (20.0\%) &    following (20.4\%) &       but (20.5\%) \\
            Digital Library &         allow (11.1\%) &              for (27.6\%) &         that (30.3\%) &      with (31.8\%) \\
            Digital Society &           for (45.8\%) &  for this reason (50.0\%) &        where (62.5\%) &    allow (100.0\%) \\
            Health          &       in this (16.7\%) &           around (20.0\%) &     based on (22.2\%) &   through (25.0\%) \\
            Infrastructure  &     following (18.2\%) &       to provide (25.0\%) &           on (41.3\%) &        in (42.8\%) \\
            Physics         &         while (33.3\%) &             from (34.8\%) &      in this (42.9\%) &       for (43.6\%) \\
            Smart City      &        within (11.1\%) &       to perform (15.4\%) &          who (15.4\%) &     where (15.8\%) \\
            Sustainability  &         within (3.8\%) &             with (13.7\%) &   to provide (14.3\%) &     while (14.3\%) \\
            Telecomm        &    to provide (16.7\%) &            which (19.4\%) &      so that (20.0\%) &     given (20.0\%) \\
            \bottomrule
        \end{tabular}}
    \end{subtable}
\end{table}

\subsection{Implications for Causality Detection and Extraction}
\label{sec:study:implications}
Based on the results of our case study, we draw the following conclusions: Causality is prevalent in requirements artifacts and therefore matters in requirements engineering, which motivates the necessity of not only an effective and reliable approach for the automatic detection and extraction of causal requirements, but also an investigation of the impact causality in requirements artifacts has. The complexity of causal relations is confined since they usually consist of a single cause and effect relationship in all observed, eligible domains. However, for an approach that aims to extract the causal relationship to be applicable in practice, it needs to comprehend also more complex relations containing at least two to three and at best an arbitrary number of causes and effects. Understanding conjunctions, disjunctions, and negations is consequently imperative to fully capture the relationships between causes and effects and ensure the applicability of a detection and extraction approach. \textit{Two-sentence causality} and \textit{event chains} occur only rarely. Thus, both aspects can initially be neglected in the development of the approaches and preserve coverage of more than 92~\% of the analyzed sentences. The dominance of \textit{explicit} over \textit{implicit} causal relations in the observed sentences simplifies the detection and extraction of causality. The information about both causes and effects is embedded directly in the sentences so that an approach requires little or no \textit{implicit} knowledge. The analysis of the precision values reveals that most of the used cue phrases are ambiguous. Consequently, automatic detection and extraction methods require a deep understanding of language as the presence of certain cue phrases is insufficient as an indicator for causality. Instead, a combination of the syntax and semantics of the sentence has to be considered to reliably detect causal relations.

\subsection{Threats to Validity}
\label{sec:study:threats}
\paragraph{Internal Validity} A threat to the internal validity is the annotation process itself as any annotation task is subjective to a certain degree. This is especially relevant for more ambiguous categories like \autour{Explicit}, as \textit{implicit} causality is difficult to determine. Two mitigation strategies were performed to minimize the bias of the annotators: First, we conducted a workshop prior to the annotation process to ensure a common understanding of causality. Second, we assessed the inter-rater agreement by using multiple metrics (Cohen's Kappa, Agreement Score, and Gwet's AC1). However, it has to be noted that all categories except \autour{Causality} are dependent on a sentence's classification regarding that category, which may imply a confounding factor for the inter-rater agreement on the other categories. This manifests in the calculation of the inter-rater agreement, where all categories except \autour{Causality} are calculated based on the 499 causal sentences. We argue however that the other categories are irrelevant for non-causal sentences as they only refer to the causal relation contained by a sentence, hence this confounding factor is deemed minimal. Apart from that, the inter-rater agreement is not domain-specific, which implies that it is not possible to identify, whether certain domains caused more disagreement among the raters. We deem the general inter-rater agreement reported in Tab.~\ref{Tab:measures} sufficient but recommend considering this aspect for replications and future studies intensifying the domain-dependent aspect of causality.
Furthermore, restricting the manual detection of causal relations to a span of a maximum of two sentences poses also a threat to internal validity, as the potential existence of causal relations that are spread across more than two sentences can neither be confirmed nor denied based on our investigation. We see this threat to be minimal as the relationship between one-sentence-causality and two-sentence-causality allows for the assumption, that the further elements of a causal relation are spread apart, the more unlikely the existence of such a causal relation is. Extrapolating from the low number of sentences categorized as two-sentence-causality gives us reason to assume that disregarding causal relations spread across three sentences or more is negligible for this initial case study.

\paragraph{External Validity} To achieve reasonable generalizability, we selected requirements documents from different domains and years. As Fig.~\ref{fig:dataset} shows, our data set covers a variety of domains, but the distribution of the sentences is imbalanced. The domains aerospace, data analytics, and smart city account for a large share in the data set (9,724 sentences), while the other 15 domains are rather underrepresented. We mitigate this threat to validity by including a domain-specific investigation reported in the scope of RQ 6, which confirms that the occurrence of causality is to a large degree domain-independent. Future studies should however expand to more documents emerging from underrepresented domains to allow a more general reflection upon different aspects of causality in requirements documents.

\section{Approach: Detecting Causal Requirements}
\label{sec:detection}
This section presents the implementation of our causal classifier. To this end, we describe a variety of applied methods followed by a report of the results of our experiments, in which we compare the performance of the individual methods and draw a conclusion in regards to applicability.

\subsection{Methods}
\label{sec:detection:methods}

\paragraph{Rule-based Approach}
Instead of using a random classifier as the baseline approach, we involve simple regex expressions for causality detection. We iterate through all sentences in the test set and check if one of the phrases listed in Tab.~\ref{Tab:cuePhrases} is contained. In the positive case, the sentence is classified as causal and vice versa. As discussed in Sect.~\ref{sec:terminology}, the classification of a sentence as causal based on the occurrence of a cue phrase -- which the baseline approach represents -- is reasonable to assume.

\paragraph{Machine Learning-based Approach}
As a second approach, we investigate the use of \textit{supervised} ML models that learn to predict causality based on the labeled data set. Specifically, we employ established binary classification algorithms: Naive Bayes (NB), Support Vector Machines (SVM), Random Forest (RF), Decision Tree (DT), Logistic Regression (LR), Ada Boost (AB), and K-Nearest Neighbor (KNN). To determine the best hyperparameters for each binary classifier, we apply Grid Search, which fits the model on every possible combination of hyperparameters and selects the most performant. We use two different methods as word embeddings: Bag of Words (BoW) and Term Frequency-Inverse Document Frequency (TF-IDF). In Tab.~\ref{Tab:experimentalResults} we report the classification results of each algorithm as well as the best combination of hyperparameters.

\paragraph{Deep Learning-based Approach}
With the rise of Deep Learning (DL), more and more researchers are using DL models for Natural Language Processing (NLP) tasks. In this context, the Bidirectional Encoder Representations from Transformers (BERT) model~\cite{devlin19} is prominent and has already been used for question answering and named entity recognition. BERT is pre-trained on large corpora and can therefore easily be fine-tuned for any downstream task without the need for much training data (Transfer Learning). In our paper, we make use of the fine-tuning mechanism of BERT and investigate to which extent it can be used for causality detection of requirement sentences. First, we tokenize each sentence. BERT requires input sequences with a fixed length (maximum 512 tokens). Therefore, for sentences that are shorter than this fixed length, padding tokens (PAD) are inserted to adjust all sentences to the same length. Other tokens, such as the classification (CLS) token, are also inserted in order to provide further information of the sentence to the model. CLS is the first token in the sequence and represents the whole sentence (i.e., it is the pooled output of all tokens of a sentence). For our classification task, we mainly use this token because it stores the information of the whole sentence. We feed the pooled information into a single-layer feedforward neural network that uses a softmax layer, which calculates the probability that a sentence is causal or not. We tune BERT in three different ways and investigate their performance:
\begin{compactitem}
  \item \textbf{BERT\textsubscript{Base}} In the base variant, the sentences are tokenized as described above and put into the classifier. To choose a suitable fixed length for our input sequences, we analyzed the lengths of the sentences in our data set. Even with a fixed length of 128 tokens, we cover more than 97~\% of the sentences. Sentences containing more tokens are shortened accordingly. Since this is only a small amount, only a little information is lost. Thus, we chose a fixed length of 128 tokens instead of the maximum possible 512 tokens to keep BERT's computational requirements to a minimum. 
  \item \textbf{BERT\textsubscript{POS}} Studies have shown that the performance of NLP models can be improved by providing explicit prior knowledge of syntactic information to the model~\cite{sundararaman2019}. Therefore, we enrich the input sequence with syntactic information and feed it into BERT. More specifically, we add the corresponding part-of-speech (POS) tag to each token by using the spaCy NLP library~\cite{spacy2}. One way to encode the input sequence with the corresponding POS tags is to concatenate each token embedding with a hot encoded vector representing the POS tag. Since the BERT token embeddings are high-dimensional, the impact of a single added feature (i.e., the POS tag) would be low. Contrary, we hypothesize that the syntactic information has a higher impact if we annotate the input sentences directly with the POS tags and then put the annotated sentences into BERT. This way of creating linguistically enriched input sequences has already proven to be promising during the development of the NLPL word embeddings~\cite{fares17}. Fig.~\ref{fig:inputSequence} shows how we incorporated the POS tags into the input sequence. By extending the input sequence, the fixed length for the BERT model has to be adapted accordingly. After further analysis, a length of 384 tokens proved to be reasonable.
  \item \textbf{BERT\textsubscript{DEP}} Similar to the previous fine-tuning approach, we follow the idea of enriching the input sequence by linguistic features. Instead of using the POS tags, we use the dependency (DEP) tags (see Fig.~\ref{fig:inputSequence}) of each token. Thus, we provide knowledge about the grammatical structure of the sentence to the classifier. We hypothesize that this knowledge has a positive effect on the model performance, as a causal relation is a specific grammatical structure (e.g., it often contains an adverbial clause) and the classifier can learn causal specific patterns in the grammatical structure of the training instances. The fixed token length was also increased to 384 tokens.
\end{compactitem}

\begin{figure}
\begin{mdframed} 
   {\scriptsize \textsc{Bert}\textsubscript{Base}: If the process fails, an error message is shown.}
   
   {\scriptsize \textsc{Bert}\textsubscript{POS}: If\_\blue{SCONJ} the\_\blue{DET} process\_\blue{NOUN} fails\_\blue{VERB} ,\_\blue{PUNCT} an\_\blue{DET} error\_\blue{NOUN} message\_\blue{NOUN} is\_\blue{AUX} shown\_\blue{VERB} .\_\blue{PUNCT}}
   
   {\scriptsize \textsc{Bert}\textsubscript{DEP}: If\_\orange{mark} the\_\orange{det} process\_\orange{nsubj} fails\_\orange{advcl} ,\_\orange{punct} an\_\orange{det} error\_\orange{compound} message\_\orange{nsubjpass} is\_\orange{auxpass} shown\_\orange{ROOT} .\_\orange{punct}}
   \end{mdframed}
   \vspace{-.4cm}
    \caption{Input sequences used for our different BERT fine-tuning models. POS tags are marked \blue{blue} and DEP tags are marked \orange{orange}.}
    \label{fig:inputSequence}
\end{figure}

\subsection{Evaluation Procedure}
\label{sec:detection:evaluation}
\begin{figure*}
    \centering
    \scalebox{1}{\includegraphics[width=\textwidth]{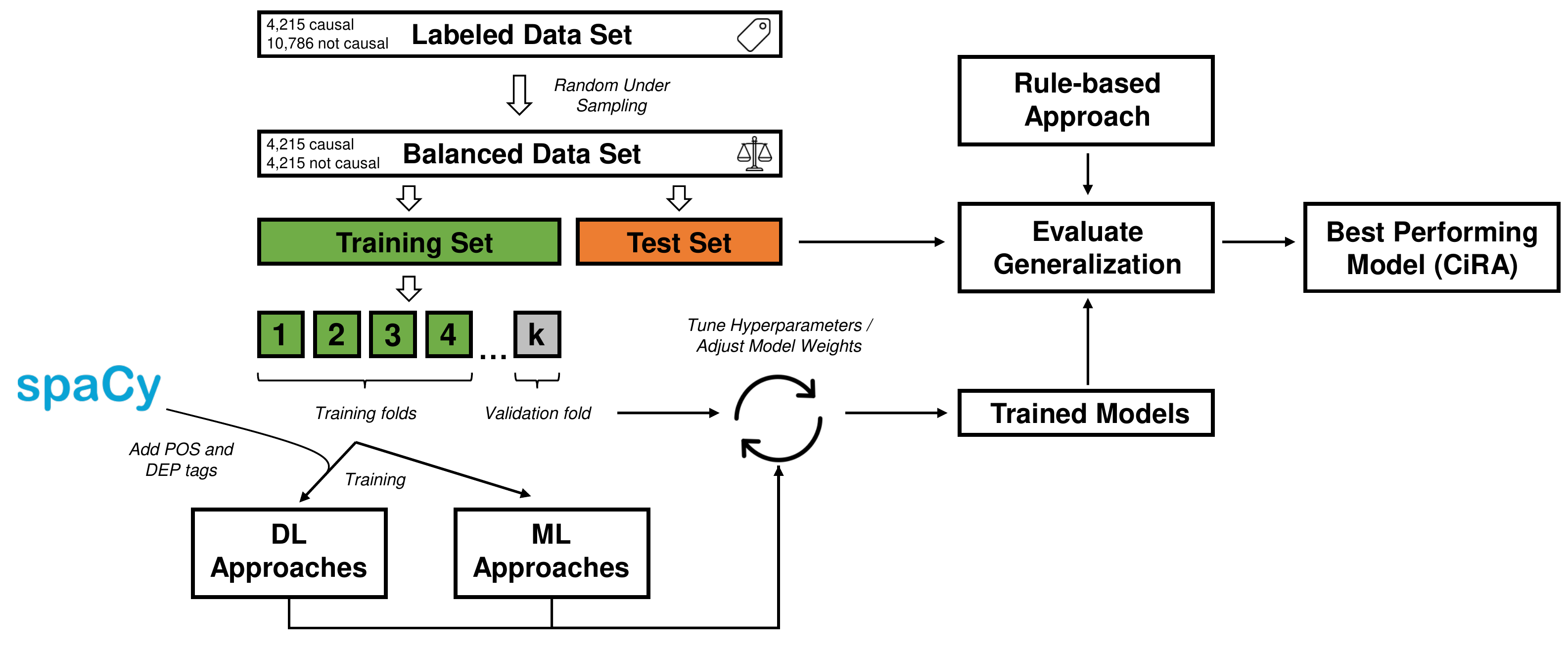}}
    \vspace{-0.6cm}
    \caption{Implementation and Evaluation Procedure of our Binary Classifier}
    \label{fig:implementationProcedure}
\end{figure*}

Our labeled data set is imbalanced as only 28.1~\% are positive samples. To avoid the class imbalance problem, we apply Random Under Sampling (see Fig.~\ref{fig:implementationProcedure}). We randomly select sentences from the majority class and exclude them from the data set until a balanced distribution is achieved. Our final data set consists of 8,430 sentences of which 4,215 are causal and the other 4,215 are non-causal. We follow the idea of cross-validation and divide the data set into a training, validation, and test set. The training set is used for fitting the algorithm, while the validation set is used to tune its parameters. The test set is utilized for the evaluation of the algorithm based on real-world unseen data. We opt for 10-fold cross-validation as a number of studies have shown that a model that has been trained this way demonstrates low bias and variance~\cite{James13}. We use standard metrics for evaluating our approaches: Accuracy, Precision, Recall, and F\textsubscript{1} score~\cite{James13}. Since a single run of a \textit{k}-fold cross-validation may result in a noisy estimate of model performance, we repeat the cross-validation procedure five times and average the scores from all repetitions (see Tab.~\ref{Tab:experimentalResults}). When interpreting the metrics, it is important to consider which misclassification (False Negative or False Positive) matters most, respectively causes the highest costs. Since causality detection is supposed to be the first step towards automatic causality extraction, we favor Recall over Precision. A high Recall corresponds to a greater degree of automation of causality extraction because it is easier for users to discard False Positives than to manually detect False Negatives. Consequently, we seek high Recall to minimize the risk of missed causal sentences and acceptable Precision to ensure that users are not overwhelmed by False Positives. 

\subsection{Experimental Results}
\label{sec:detection:results}
Tab.~\ref{Tab:experimentalResults} demonstrates the inability of the rule-based baseline approach to distinguish between causal (F\textsubscript{1} score: 66~\%) and non-casual (F\textsubscript{1} score: 64~\%) sentences. This coincides with our observation from the first case study that classifying sentences as causal or non-causal based on the occurrence of cue phrases is not suitable for causality detection. In comparison, most ML-based approaches (except KNN and DT) show a better performance. The best performance in this category is achieved by RF with an Accuracy of 78~\% (gain of 13~\% compared to baseline approach). The overall best classification results are achieved by our DL-based approaches. All three variants were trained with the hyperparameters recommended by Devlin et al.~\cite{devlin19}. Even the vanilla \textbf{BERT\textsubscript{Base}} model shows a great performance in both classes (F\textsubscript{1} score $\geq$ 80~\% for causal and non-causal). Interestingly, enriching the input sequences with syntactic information did not result in a significant performance boost. \textbf{BERT\textsubscript{POS}} even has a slightly worse Accuracy value of 78~\% (difference of 2~\% compared to \textbf{BERT\textsubscript{Base}}). An improvement of the performance can be observed in the case of \textbf{BERT\textsubscript{DEP}}, which has the best F\textsubscript{1} score for both classes among all the other approaches and also achieves the highest Accuracy value of 82~\%. Compared to the rule-based and ML-based approaches, \textbf{BERT\textsubscript{DEP}} yields an average gain of 11.06~\% in macro-Recall and 11.43~\% in macro-Precision. Interesting is a comparison with \textbf{BERT\textsubscript{Base}}. \textbf{BERT\textsubscript{DEP}} shows better values across all metrics, but the difference is only marginal. This indicates that \textbf{BERT\textsubscript{Base}} already has a deep language understanding due to its extensive pre-training and therefore can be tuned well for causality detection without much further input. However, over all five runs, the use of the dependency tags shows a small but not negligible performance gain - especially regarding our main decision criterion: the Recall value (85~\% for causal and 79~\% for non-causal). Therefore, we choose \textbf{BERT\textsubscript{DEP}} as our final approach (CiRA).

\begin{table}
    \centering
    \caption{Recall, Precision, F\textsubscript{1} scores (per class) and Accuracy. We report the averaged scores over five repetitions. Best results for each metric are highlighted in \textbf{bold}.}
     \vspace{-0.2cm}
    \label{Tab:experimentalResults}
    \resizebox{\textwidth}{!}{\begin{tabular}{lllccccccc} 
        \toprule
        & & & \multicolumn{3}{c}{\textbf{Causal (Support: 435)}} & \multicolumn{3}{c}{\textbf{Not Causal (Support: 408)}} & \multicolumn{1}{l}{} \\
        & & Best hyperparameters & Recall & Precision & F1 & Recall & Precision & F1 & Accuracy \\
        
        \textbf{Rule based} &  & - & 0.65 & 0.66 & 0.66 & 0.65 & 0.63 & 0.64 & 0.65 \\
        \hline
        
        \multirow{7}{*}[-5em]{\textbf{ML based}} & NB  & \begin{tabular}{@{}l@{}}alpha: 1, fit\_prior: True,\\embed: BoW\end{tabular} & 0.71 & 0.7 & 0.71 & 0.68 & 0.69 & 0.69 & 0.7 \\  
        \addlinespace
        & SVM & \begin{tabular}{@{}l@{}}C: 50, gamma: 0.001,\\kernel: rbf, embed: BoW\end{tabular} & 0.68 & 0.8 & 0.73 & 0.82 & 0.71 & 0.76 & 0.75 \\  
        \addlinespace
        & RF & \begin{tabular}{@{}l@{}}criterion: entropy, max\_features: auto,\\ n\_estimators: 500, embed: BoW\end{tabular} & 0.72 & \textbf{0.82} & 0.77  & \textbf{0.84} & 0.74 & 0.79 & 0.78 \\
        \addlinespace
        & DT  & \begin{tabular}{@{}l@{}}criterion: gini, max\_features: auto,\\ splitter: random, embed: TF-IDF \end{tabular} & 0.65 & 0.68 & 0.66 & 0.67 & 0.65 & 0.66 & 0.66 \\
        \addlinespace
        & LR  & \begin{tabular}{@{}l@{}} C: 1, solver: liblinear,\\  embed: TF-IDF\end{tabular} & 0.71 & 0.78 & 0.74  & 0.79 & 0.72  & 0.75 & 0.75 \\
        \addlinespace
        & AB & \begin{tabular}{@{}l@{}}algorithm: SAMME.R, n\_estimators: 200,\\ embed: BoW \end{tabular} & 0.67 & 0.78  & 0.72 & 0.8 & 0.7  & 0.75 & 0.74 \\
        \addlinespace
        & KNN & \begin{tabular}{@{}l@{}}algorithm: ball\_tree, n\_neighbors: 20,\\ weights: distance, embed: TF-IDF \end{tabular} & 0.61 & 0.68 & 0.64 & 0.7 & 0.63 & 0.66 & 0.65 \\ 
        \hline
        
        \multirow{3}{*}{\textbf{DL based}} & BERT\textsubscript{Base} & \multirow{3}{5cm}{batch\_size: 16, learning\_rate: 2e-05, weight\_decay: 0.01, optimizer: AdamW} & 0.83 & 0.80 & 0.82 & 0.78 & 0.82 & 0.80 & 0.81 \\
        & BERT\textsubscript{POS} & & 0.82 & 0.76 & 0.79 & 0.71 & 0.83 & 0.77 & 0.78 \\
        & BERT\textsubscript{DEP} (CiRA) &  & \textbf{0.85} & 0.81 & \textbf{0.83} & 0.79 & \textbf{0.84} & \textbf{0.81} & \textbf{0.82} \\
        \bottomrule
    \end{tabular}}
\end{table}

\section{Case Study 2: Effects of Causality}
\label{sec:effects}

After discussing first empirical evidence on the extent and complexity to which causality is used in NL requirements in our first case study (C1) and constructing a reasonably effective approach for automatic causality detection (C2), we aim to corroborate the relevance of causality for requirements in a second case study (C3). Here, we investigate the impact of causal relations on the features of requirements, where we consider features to be observable attributes of individual requirements (e.g., their lead-time). This investigation emerges from an ongoing academia-industry collaboration in a larger context. Our exploratory case study has two goals: first, we aim to demonstrate an independent use case of the automatic causality detection approach. While automatic causality detection as presented in Sect.~\ref{sec:detection} can be used as a precursor to automatic causality extraction and therefore as one step in a pipeline towards automatic test case generation, we explore considering the occurrence of causality as an aspect of requirements quality and consequently automatic causality detection as a metric to estimate requirements quality. An effective tool-supported approach for detecting causality in NL requirements allows exploring the eligibility of causality as an aspect of requirements quality. Gathering the first empirical evidence towards this is the second goal of this exploratory case study. First empirical evidence can be gathered by investigating the correlation between the occurrence of causality in requirements and features of these requirements.

\subsection{Research Questions}
\label{sec:effects:rqs}

We are interested in the impact that the occurrence of causal relations in natural language requirements has on important features of these requirements. Empirical evidence for an impact of causality on these features would allow the assumption that the use of causality in a NL requirement contributes to the requirement's quality. While a definite connection cannot be determined based on a correlation analysis alone, this exploratory case study rather aims towards providing first insights into the feasibility of using causality as a quality aspect for requirements and opening up a more detailed discussion regarding specific features. We select the following features for our analysis:
\begin{compactitem}
    \item \textbf{Lead-Time}: the time from the inception until the completion of a requirement.
    \item \textbf{Consolidated state}: the type of final state in which the requirement ends in.
    \item \textbf{Volatility}: number of state changes which the requirement undergoes.
\end{compactitem}
The selection of attributes is inspired by research with comparable objectives, which used lead-time and the resulting consolidated state~\cite{olsson2019empirical} as well as the volatility~\cite{wnuk2015supporting} of requirements as dependent variables to estimate the impact of requirements attributes on the downstream development process. A data set eligible for this evaluation needs to provide information in form of a state log, where each entry in the log documents the author, timestamp, and state code. The state codes represent the different states which a requirement traverses during its life cycle from its inception to its completion. The lead-time consequently constitutes the time delta between the first and the last state log entry. The consolidated state is the final state of the state log. The volatility denotes the number of entries in the state log, as it directly correlates with the number of additional development cycles the requirement has to traverse (e.g., by being pushed back to earlier states when repeating one development cycle). We want to investigate whether a statistically significant difference can be determined in the distribution of lead-time, consolidated state, and volatility between requirements that use causal relations versus requirements that do not. To this extent, we aim at providing answers to the following research questions (RQ):
\begin{compactitem}
    \item \textbf{RQ 7}: Does the use of causality in an NL requirement correlate with its lead-time?
    \item \textbf{RQ 8}: Does the use of causality in an NL requirement correlate with its consolidated state?
    \item \textbf{RQ 9}: Does the use of causality in an NL requirement correlate with its volatility?
\end{compactitem}
These attributes have been chosen as an eligible representation of the requirements' comprehensibility and degree of ambiguity. As elaborated in earlier sections, we hypothesize that the clear semantic structure of a causal relation promotes comprehensibility and mitigates the ambiguity of requirements. This would result in shorter lead times, a greater likelihood of a successful outcome, and less volatility, as the requirement has to undergo fewer iterations in the development life-cycle.

\begin{table}[]
\caption{Features of the data set}
\vspace{-0.2cm}
    \label{tab:datasetfeatures}
    \centering
    \begin{tabular}{|p{1.7cm}|p{7cm}|l|} \hline
        \textbf{Feature} & \textbf{Description} & \textbf{Datatype} \\ \hline
        ID & Unique identifier of the requirement & numeric \\ 
        Description & Textual description containing a varying amount of NL sentences & text \\
        State log & History of state changes & categorical list \\
        Date of creation & Inception date of the requirement & date \\
        \hline 
    \end{tabular}
\end{table}

\subsection{Study Design}
\label{sec:effects:design}

\paragraph{Study Objects}
The study is performed on a data set for an industrial, proprietary project. The owning, multi-national case company develops and globally distributes software-intensive products for a B2C market. The number of engineers involved with the product line of the data set in question varied from 1000 to 4000 worldwide. The original data set, pre-processed by Olsson et al.~\cite{olsson2019empirical} contains 4446 requirements collected in 2016\footnote{The proprietary data set cannot be disclosed at the time of submission as it contains critical company information.}. The data set has been chosen because it contains the aforementioned features necessary for the evaluation, other than the data sets used for the first case study: most of the requirements in the data set contain a state log documenting the requirement's life cycle from its inception as a \textit{New Feature} (NF) to its final state as either \textit{Execution completed} (EC) or \textit{Discarded} (D). The newly created requirements undergo the initial triage in a state called M0. Next, upon considered viable and sufficiently justified, the requirement candidates are prioritized in a project prioritization state (similar to backlog prioritization), called M1. Finally, the prioritized requirements are hand-shaken with the developer teams in a state called M2~\cite{fricker2010handshaking}. When a requirement is unclear at the M2 state, it is pushed back to M1 for re-prioritization. Similarly, a requirement is pushed back to M0 when questions and uncertainties arise during requirements prioritization. These backward transitions unusually increase the lead-time. The features relevant for this study are described in Tab.~\ref{tab:datasetfeatures}. Further attributes were generated based on the existing features:
\begin{compactitem}
    \item \textbf{Sentences}: The number of sentences occurring in the \textit{Description} field is counted via sentence tokenization.
    \item \textbf{Causal Relations}: The number of causal sentences in the \textit{Description} field is counted by applying the CiRA-tool presented in Sect.~\ref{sec:detection} to each sentence.
    %\item \textbf{Causal Percentage}: The fraction of causal sentences used in a requirement is calculated by dividing \textit{Causal Relations} by \textit{Sentences}.
    \item \textbf{Lead-Time}: The lead-time is calculated as the time frame between the first and the last entry of the \textit{state log}.
    \item \textbf{Volatility}: Number of decisions counted as the number of entries in the \textit{state log}
\end{compactitem}
The data set of 4446 requirements was further pre-processed to serve the application in this second case study. Three additional filters have been applied: (1) requirements, for which the state log did not exist, were discarded. (2) Requirements with a state log authored by exactly one, specific individual, were discarded. The entries to these state logs were due to database migrations and do not contain actual information on the requirements life cycle. (3) Requirements with only one entry in the state log were discarded. These requirements do not allow calculating a meaningful lead-time. In total, 815 requirements were discarded due to the pre-processing, leaving 3631 requirements for the analysis. Tab.~\ref{tab:datasetcleaned} lists further details on the filtering process.
%(4) Entries in the state log, which did not conform the constrained state transition graph, were discarded. These invalid transitions occurred due to database migrations and were removed to not inflate the volatility of requirements.

\begin{table}[]
    \caption{Pre-processing steps}
    \vspace{-0.2cm}
    \label{tab:datasetcleaned}
    \centering
    \begin{tabular}{|c|l|r|r|} \hline
        \textbf{ID} & \textbf{Filter} & \textbf{Removed} & \textbf{Remaining} \\ \hline
        1 & Missing consolidated state log & 176 & 4270 \\
        2 & Specific, invalid author & 185 & 4085 \\
        3 & Singular state log entry & 454 & 3631 \\ \hline
        & \textbf{Total} & \textbf{815} & \textbf{3631} \\ \hline
    \end{tabular}
\end{table}

\paragraph{Data Analysis}
The research questions can be translated into statistically verifiable (i.e., refutable) hypotheses. We therefore formulate the following null hypotheses:

\begin{compactitem}
    \item \textbf{H\textsubscript{1\textsubscript{0}}}: Requirements containing different amounts of causality have the same distribution of lead-time.
    \item \textbf{H\textsubscript{2\textsubscript{0}}}: Requirements containing different amounts of causality have the same distribution of consolidated states.
    \item \textbf{H\textsubscript{3\textsubscript{0}}}: Requirements containing different amounts of causality have the same distribution of volatility.
\end{compactitem}
In all hypotheses the input variable \textit{contained amounts of causality} is tested on two levels of granularity: (G1) binary (containing at least one causal sentence vs. containing no causal sentence) and (G2) in batches (ranges of number of causal sentences). Furthermore, granularity G1 is extended to the third level of granularity (G3) where the data set is split into three subsets containing requirements of different sentence sizes. The distribution of causal sentences according to the different levels of granularity is given in Tab.~\ref{tab:granularity1&2} and Tab.~\ref{tab:granularity1&3}. Where the binary granularity G1 serves to investigate the general effect of causality and the batch granularity G2 refines this relation, the extended granularity G3 normalizes the effect according to requirement size. All hypotheses are reported using descriptive statistics and evaluated using inferential statistics. The hypothesis of independence is calculated using the Mann-Whitney U test on the binary level of granularity (G1 and G3) for the interval scale variable of lead-time and volatility, and using the Chi-square test for the categorical variable of consolidated state. For batch granularity G2 the Kruskal-Wallis test is used~\cite{friedman2017elements}. All statistical tests of independence are evaluated with a significance level $\alpha=0.05$. Where a statistical tests suggests to reject the null hypothesis of independence, the effect size of the correlation is quantified using Cohen's d for binary granularities G1 and G3~\cite{sullivan2012using} and Eta-squared for batch granularity G2~\cite{cohen1973eta}. These measures allow categorizing the magnitude of the correlation effect.

\begin{table}[]
    \caption{Distribution of sentences according to granularity G1 and G2}
    \label{tab:granularity1&2}
    \centering
    \begin{tabular}{|r|r|r|} \hline
        & \textbf{n\textsubscript{causal}} & \textbf{n\textsubscript{req}}\\ \hline
        non-causal & [0] & 1000 \\ \hline
        \multirow{7}{*}{causal} & [1, 3] & 2059 \\
        & [4, 6] & 457 \\ 
        & [7, 9] & 96 \\ 
        & [10, 12] & 15 \\ 
        & [13, 15] & 2 \\ 
        & [16, 18] & 1 \\ 
        & [19, 21] & 1 \\ \hline
        total & [0, 21] & 3631 \\  \hline
    \end{tabular}
\end{table}

\begin{table}[]
    \caption{Distribution of sentences according to granularity G1 and G3}
    \label{tab:granularity1&3}
    \centering
    \begin{tabular}{|r|rrr|r|}  \hline
         n\textsubscript{sentences} & [1, 3] & [4, 7] & [8, max] & [1,max] \\ \hline
         causal & 542 & 933 & 1156 & 2631 \\
         non-causal & 514 & 331 & 175 & 1000 \\ \hline
         total & 1056 & 1244 & 1331 & 3631 \\  \hline
    \end{tabular}
\end{table}

\subsection{Study Results}
\label{sec:effects:results}

All study results are reported in Tab.~\ref{tab:cs2_results} and explained in further detail in the following Sections.

\begin{table}[]
    \caption{P- and Cohen's d value for evaluating the respective null-hypothesis with the given granularity. Cells prefixed with * indicate where the null-hypothesis has been rejected (given significance level $\alpha=0.05$).}
    \label{tab:cs2_results}
    \centering
    \begin{tabular}{ll|r|r|r|r|r} \toprule
        \multirow{2}{*}{\textbf{Hypothesis}} & \multirow{2}{*}{\textbf{Measure}} & \multirow{2}{*}{\textbf{G1}} & \multirow{2}{*}{\textbf{G2}} & \multicolumn{3}{c}{\textbf{G3}} \\
        & & & & [1, 3] & [4, 7] & [8, max] \\ \toprule
        \multirow{2}{*}{\textbf{H\textsubscript{1\textsubscript{0}}}} & p-value & *0.0004 & *0.0082 & 0.3434 & *0.0008 & *0.0084 \\
        & effect size & 0.0514 & 0.0010 & - & 0.1181 & 0.1496 \\ \midrule
        \multirow{2}{*}{\textbf{H\textsubscript{2\textsubscript{0}}}} & p-value & 0.8497 & 0.0668 & 0.1368 & 0.1456 & 0.7661 \\
        & effect size & - & - & - & - & - \\ \midrule
        \multirow{2}{*}{\textbf{H\textsubscript{3\textsubscript{0}}}} & p-value & *0.0105 & 0.0856 & 0.2184 & *0.0011 & 0.1261 \\ 
        & effect size & 0.0742 & - & - & 0.1843 & - \\ \bottomrule
    \end{tabular}
\end{table}

\begin{figure}
    \centering
    \begin{subfigure}{.49\textwidth}
        \includegraphics[width=\linewidth]{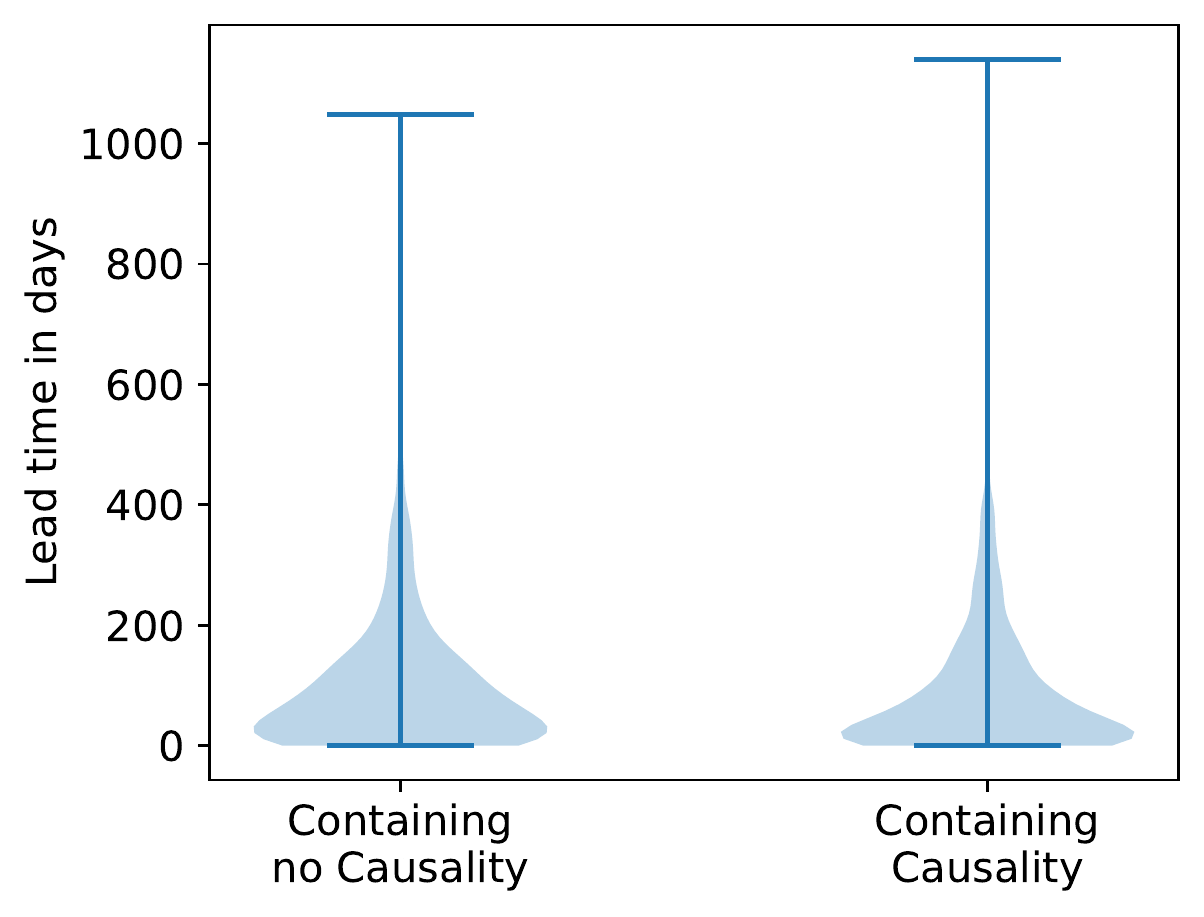}
        \caption{Binary granularity G1}
        \label{fig:cs2:h1g1}
    \end{subfigure}
    \begin{subfigure}{.49\textwidth}
        \includegraphics[width=\linewidth]{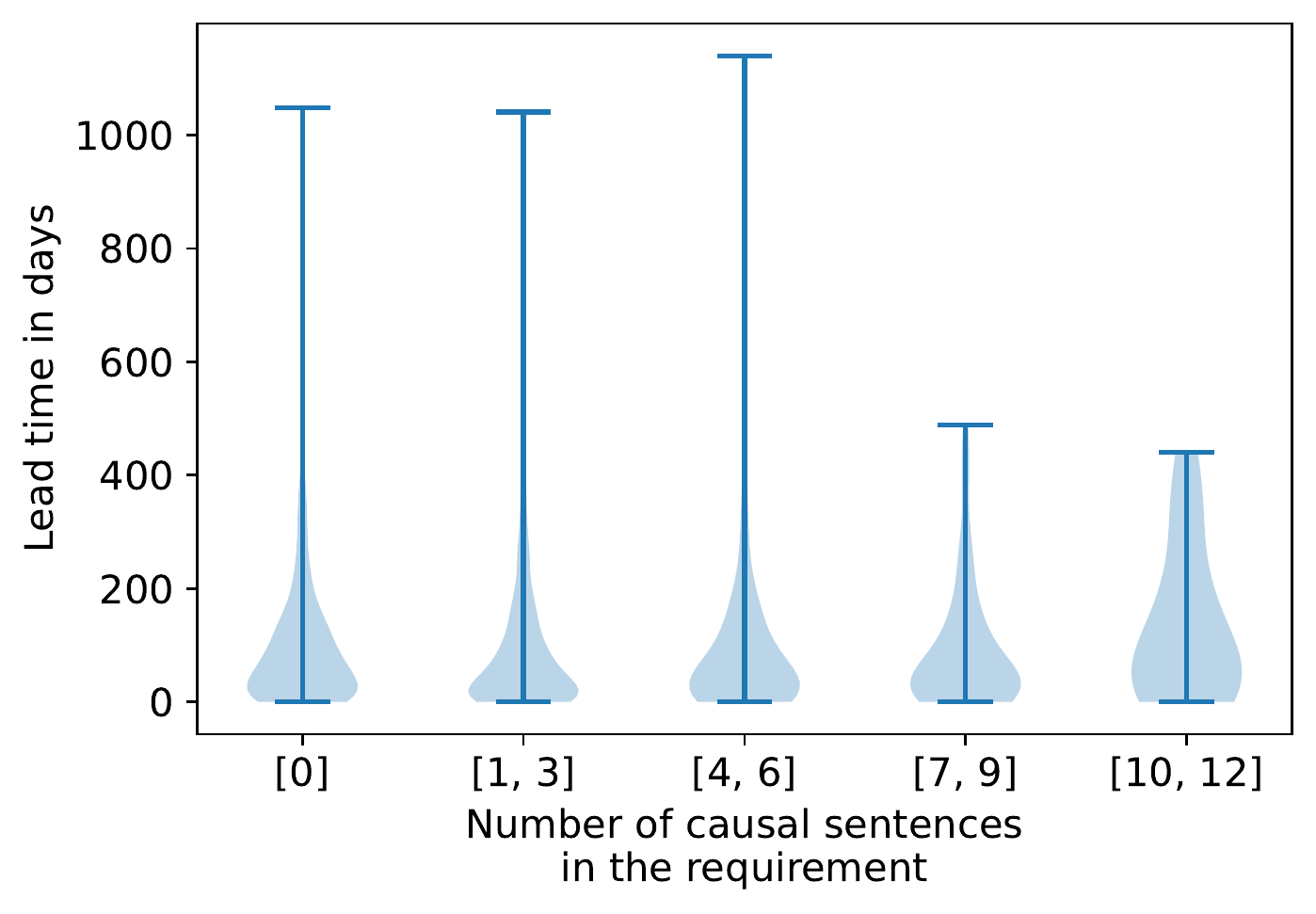}
        \caption{Batch granularity G2}
        \label{fig:cs2:h1g2}
    \end{subfigure}
    \caption{Distribution of lead-time (H\textsubscript{1\textsubscript{0}})}
    \label{fig:cs2:h1}
\end{figure}

\begin{figure}
    \centering
    \begin{subfigure}{.32\textwidth}
        \includegraphics[width=\linewidth]{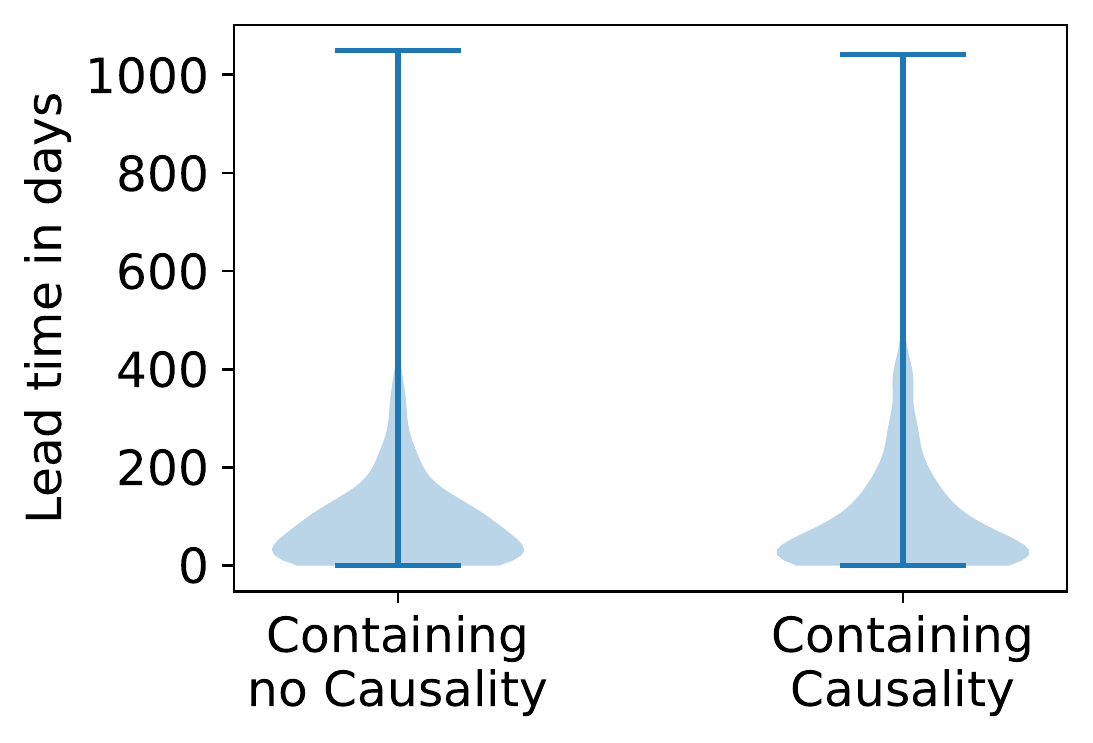}
        \caption{Small requirements}
        \label{fig:cs2:h1g3-small}
    \end{subfigure}
    \begin{subfigure}{.32\textwidth}
        \includegraphics[width=\linewidth]{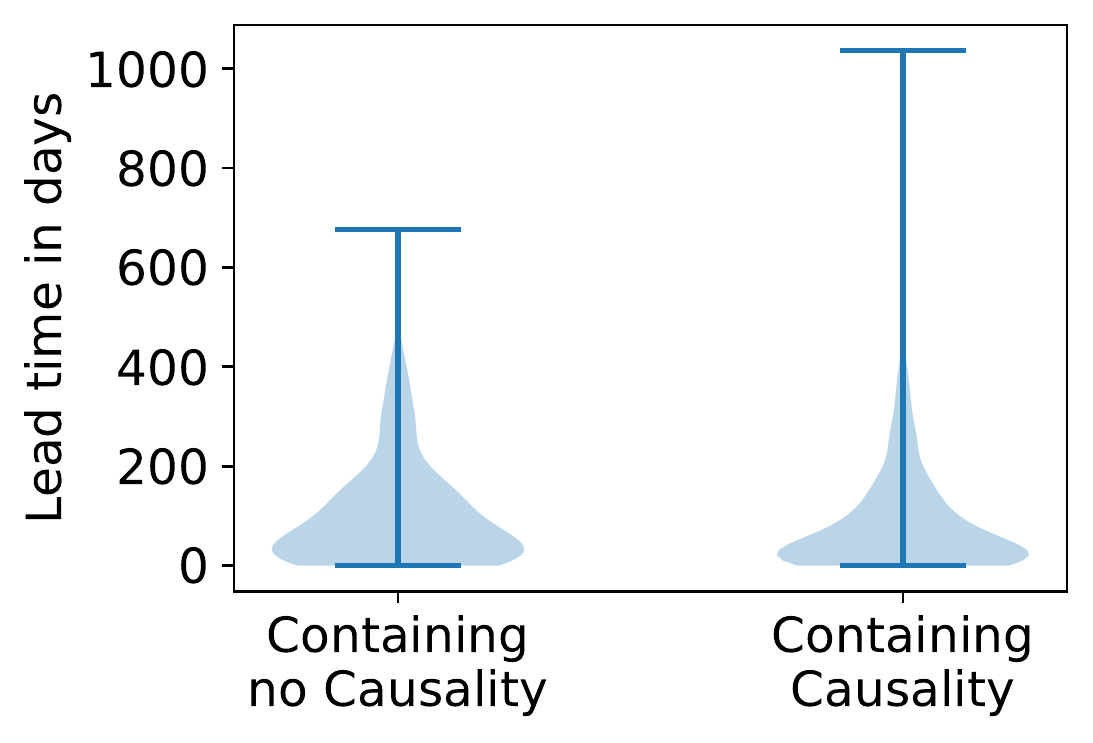}
        \caption{Medium requirements}
        \label{fig:cs2:h1g3-medium}
    \end{subfigure}
    \begin{subfigure}{.32\textwidth}
        \includegraphics[width=\linewidth]{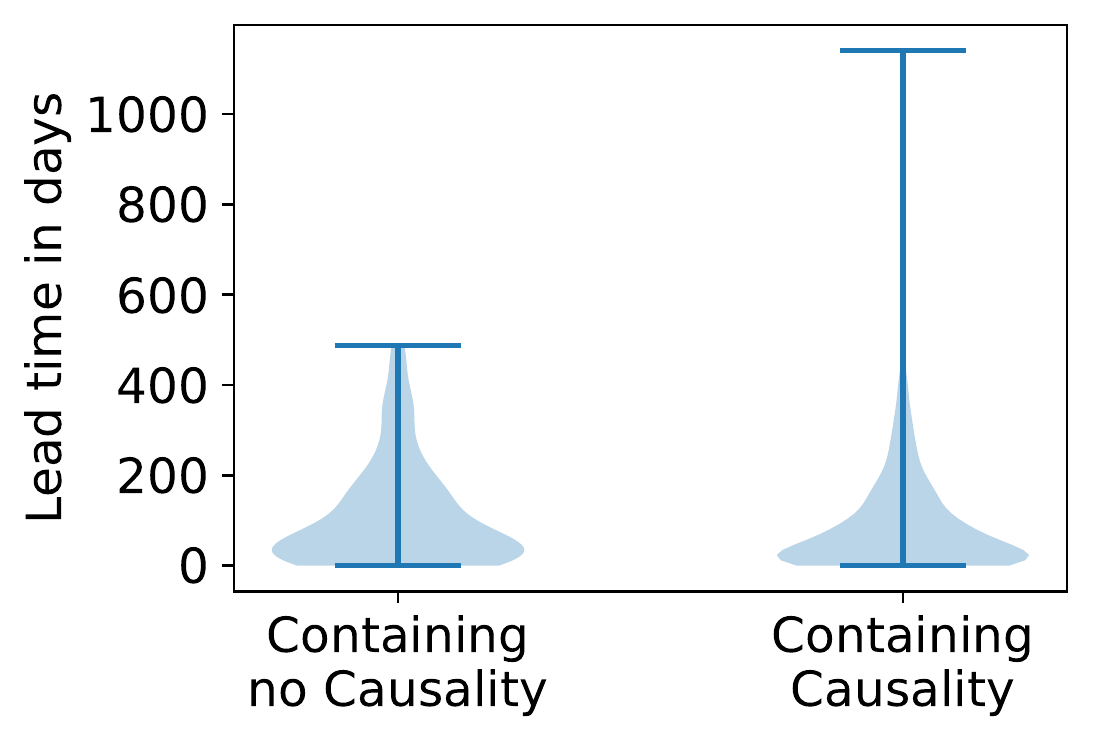}
        \caption{Large requirements}
        \label{fig:cs2:h1g3-large}
    \end{subfigure}
    \caption{Distribution of lead-time for binary granularity split by the size of requirements (H\textsubscript{1\textsubscript{0}})}
    \label{fig:cs2:h1g3}
\end{figure}

\paragraph{Correlation between causality and lead-time}
Fig.~\ref{fig:cs2:h1g1} displays the distribution of lead-time in the two binary groups in the form of violin plots and indicates that the lead-time of requirements containing at least one causal sentence is on average lower than the lead-time of requirements without any causality. The Mann-Whitney U test of independence yields a p-value of $0.00038$ far below the significance level $\alpha=0.05$, rejecting the null hypothesis of similar distribution and corroborating the found difference. Cohen's d quantifying the effect size yields 0.0514, which categorizes the effect size of the correlation as \textit{small}. For granularity G2 only batches containing more than 10 sentences were included, which leads to discarding all batches containing more than 12 causal sentences due to statistical insignificance. The results of evaluating H\textsubscript{1\textsubscript{0}} on a finer level of granularity G2 shows that the increased usage of causal sentences has a positive effect on reducing the average lead-time up until the point of using 10 or more causal sentences as shown in Fig.~\ref{fig:cs2:h1g2}. The Kruskal-Wallis test of independence suggests to reject the null hypothesis of similar distribution with a p-value of $0.0082$, but the Eta-squared value of 0.0010 categorizes the correlation as \textit{negligible}~\cite{cohen1973eta}. Evaluating the difference in distribution on granularity G3 reveals that the size of the requirement is a contributing factor to the correlation between the occurrence of causality and the lead-time: while the use of causality in small requirements has a negative effect on the lead-time, the opposite is observable for medium and large requirements, as illustrated in Fig.~\ref{fig:cs2:h1g3}. The null hypothesis of independence is accepted for small requirements with $p=0.34$ and rejected for medium and large requirements with $p=0.0008$ and $p=0.008$ respectively. The effect sizes are 0.12 and 0.15 based on the Cohen's d measure.

\begin{figure}
    \centering
    \begin{subfigure}{.49\textwidth}
        \includegraphics[width=\linewidth]{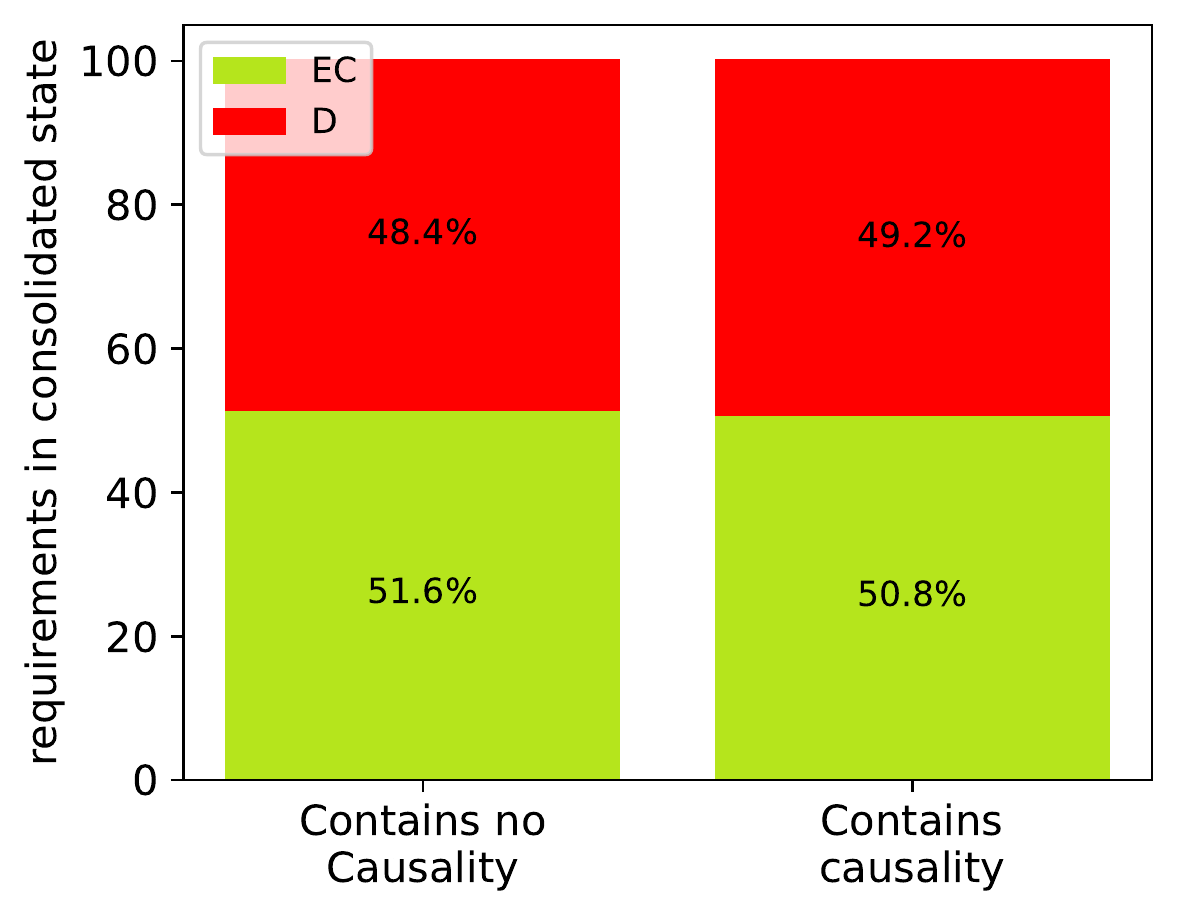}
        \caption{Binary granularity G1}
        \label{fig:cs2:h2g1}
    \end{subfigure}
    \begin{subfigure}{.49\textwidth}
        \includegraphics[width=\linewidth]{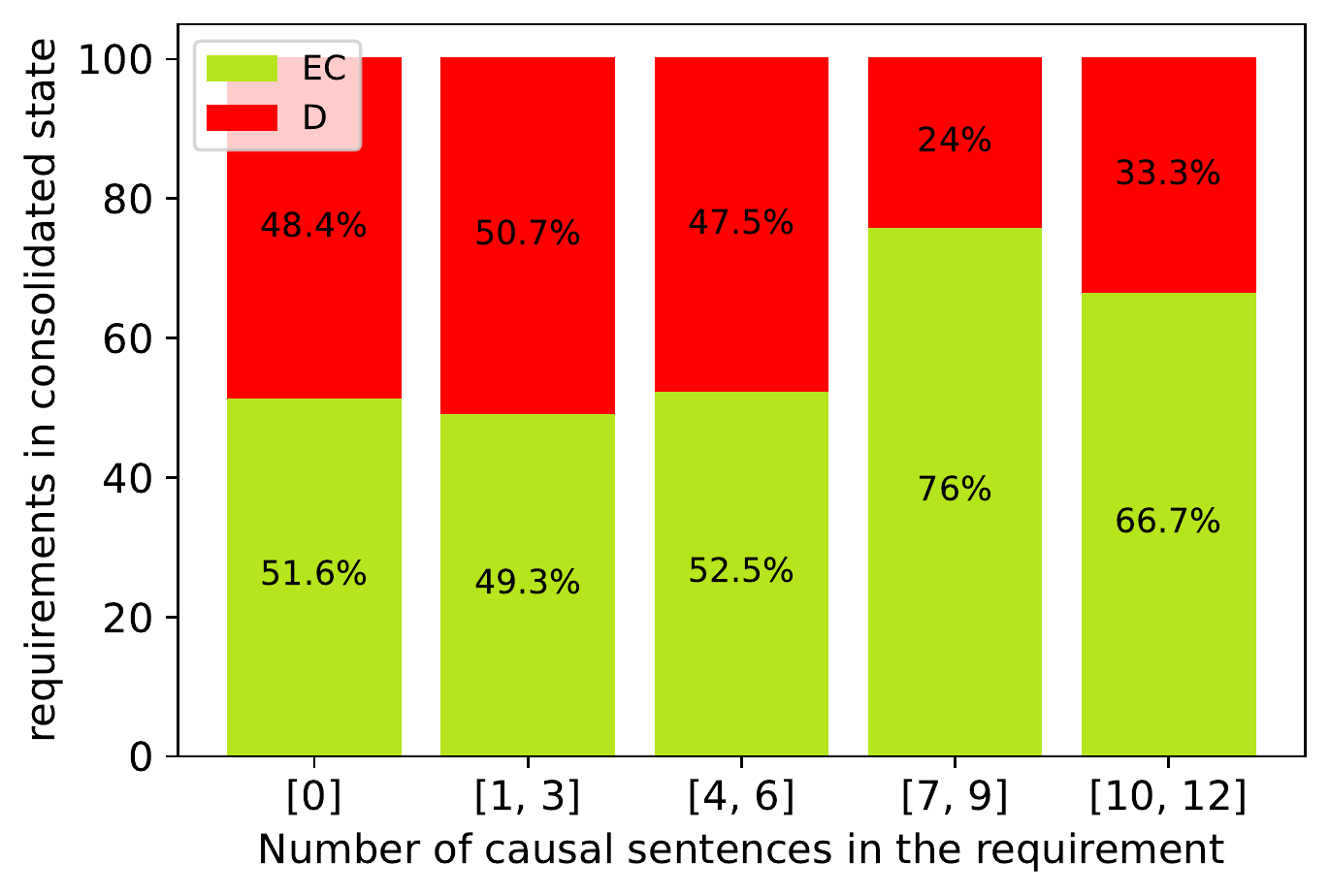}
        \caption{Batch granularity G2}
        \label{fig:cs2:h2g2}
    \end{subfigure}
    \caption{Distribution of filtered consolidated states (H\textsubscript{2\textsubscript{0}})}
    \label{fig:cs2:h2}
\end{figure}

\begin{figure}
    \centering
    \begin{subfigure}{.32\textwidth}
        \includegraphics[width=\linewidth]{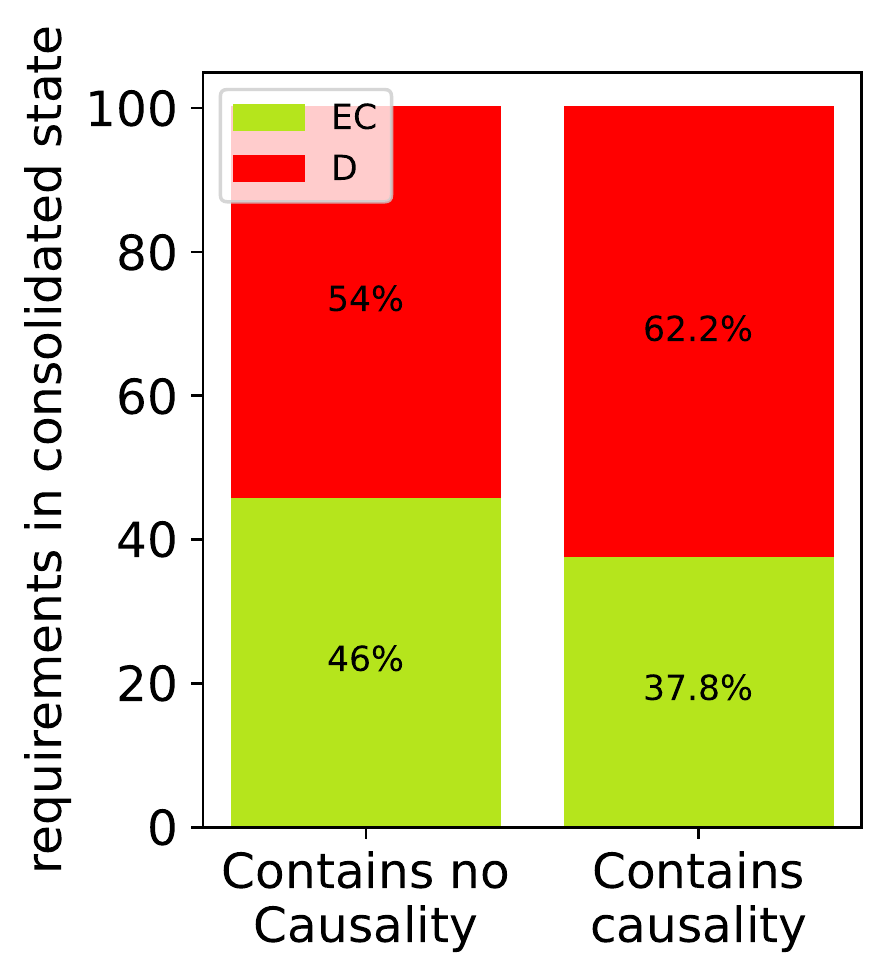}
        \caption{Small requirements}
        \label{fig:cs2:h2g3-small}
    \end{subfigure}
    \begin{subfigure}{.32\textwidth}
        \includegraphics[width=\linewidth]{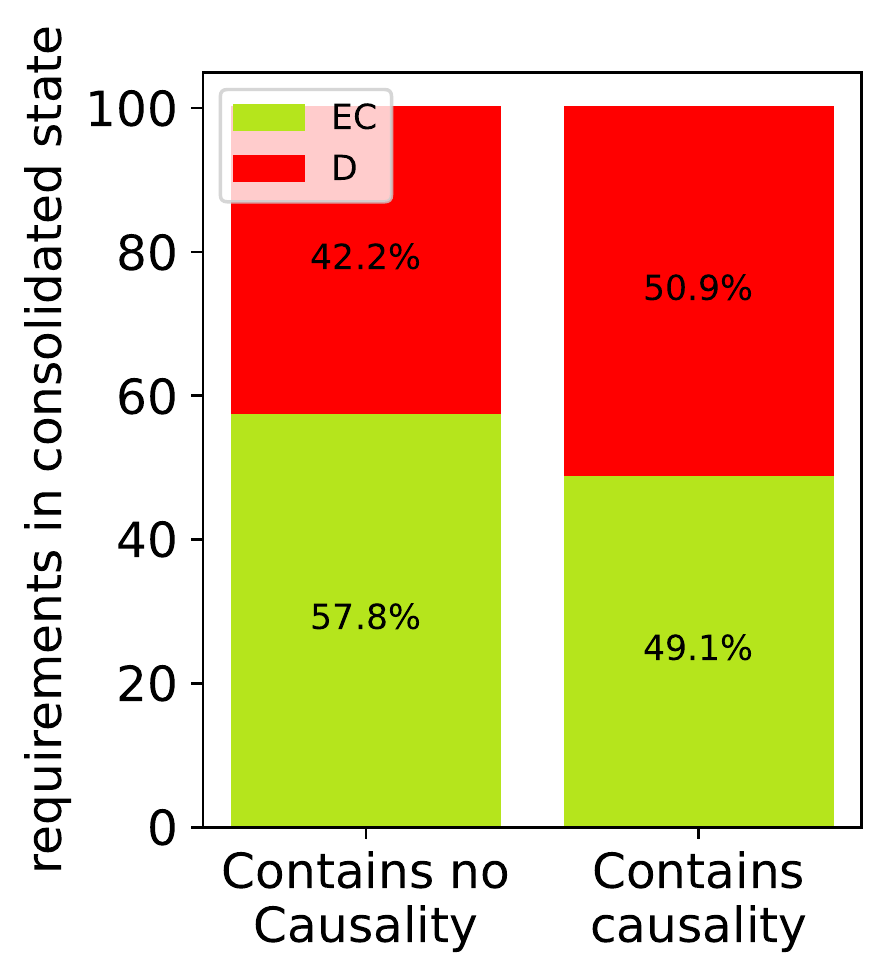}
        \caption{Medium requirements}
        \label{fig:cs2:h2g3-medium}
    \end{subfigure}
    \begin{subfigure}{.32\textwidth}
        \includegraphics[width=\linewidth]{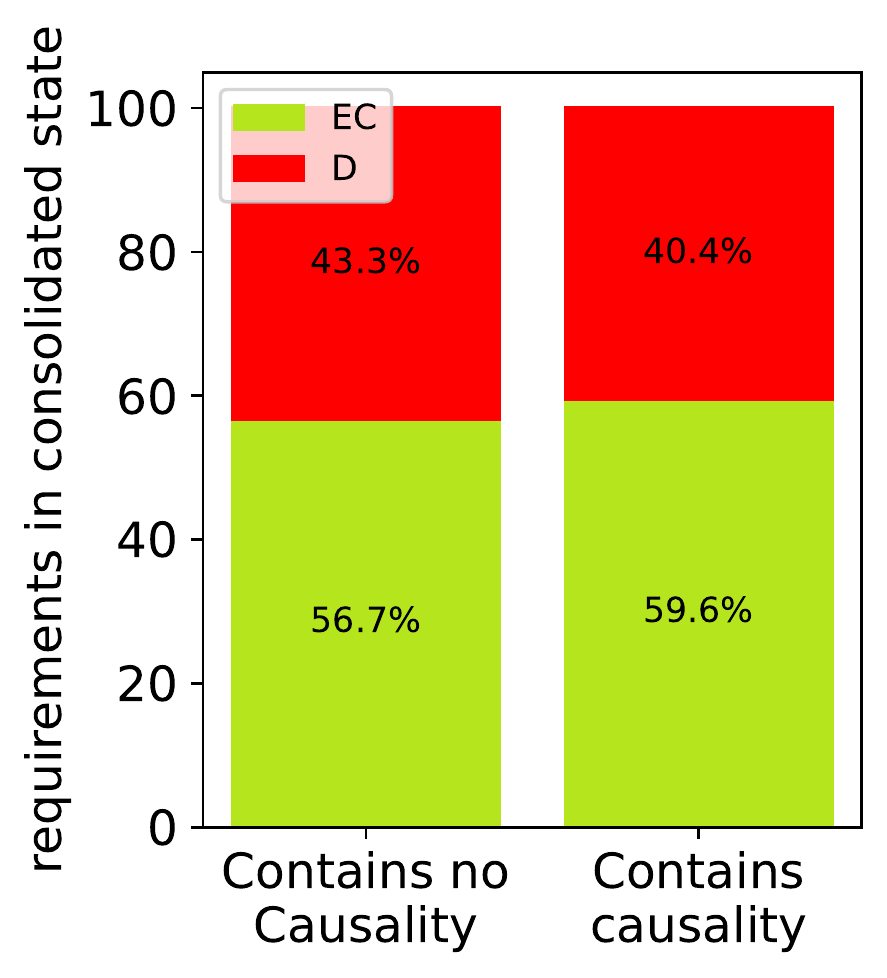}
        \caption{large requirements}
        \label{fig:cs2:h2g3-large}
    \end{subfigure}
    \caption{Distribution of filtered consolidated for binary granularity split by the size of requirements (H\textsubscript{2\textsubscript{0}})}
    \label{fig:cs2:h2g3}
\end{figure}

\paragraph{Correlation between causality and consolidated state}
The data set uses 20 categories for the variable \textit{consolidated state}, of which most represent intermediate states. The data set of 3442 requirements is filtered for all requirements in final states, which are \textit{execution completed} (EC) and \textit{discarded} (D), respectively positive and negative final state. Only the 1157 requirements in these two final states (EC: 591, D: 566) were considered for this evaluation. Fig.~\ref{fig:cs2:h2g1} illustrates the distribution of consolidated states on binary granularity G1. The Chi-square test of independence yields a p-value of 0.85 and does therefore not allow to reject the null hypothesis. Increasing the granularity to batches as displayed in Fig.~\ref{fig:cs2:h2g2} suggests a positive trend in the correlation between the occurrence of causality and a successful consolidated state, but the Kruskal-Wallis test does not allow to reject the null hypothesis of similar distribution with a p-value of $0.067$. At granularity G3, the consolidated states of requirements of different sizes correlate negatively with the occurrence of causality for small and medium requirements with only a slight positive correlation for large requirements, as seen in Fig.~\ref{fig:cs2:h2g3}. The null hypothesis of independence can however not be rejected with p-values of $0.14$, $0.15$, and $0.77$.

\begin{figure}
    \centering
    \begin{subfigure}{.49\textwidth}
        \includegraphics[width=\linewidth]{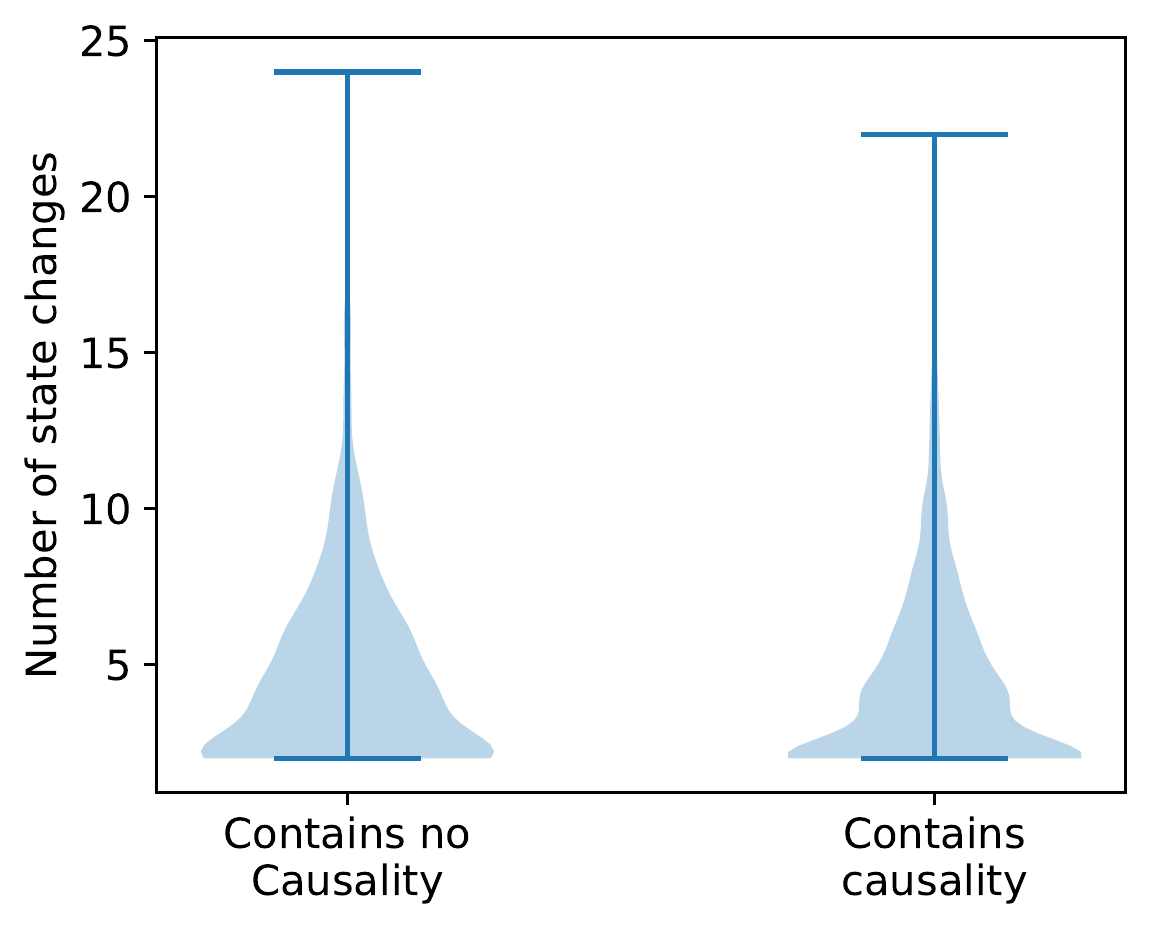}
        \caption{Binary granularity G1}
        \label{fig:cs2:h3g1}
    \end{subfigure}
    \begin{subfigure}{.49\textwidth}
        \includegraphics[width=\linewidth]{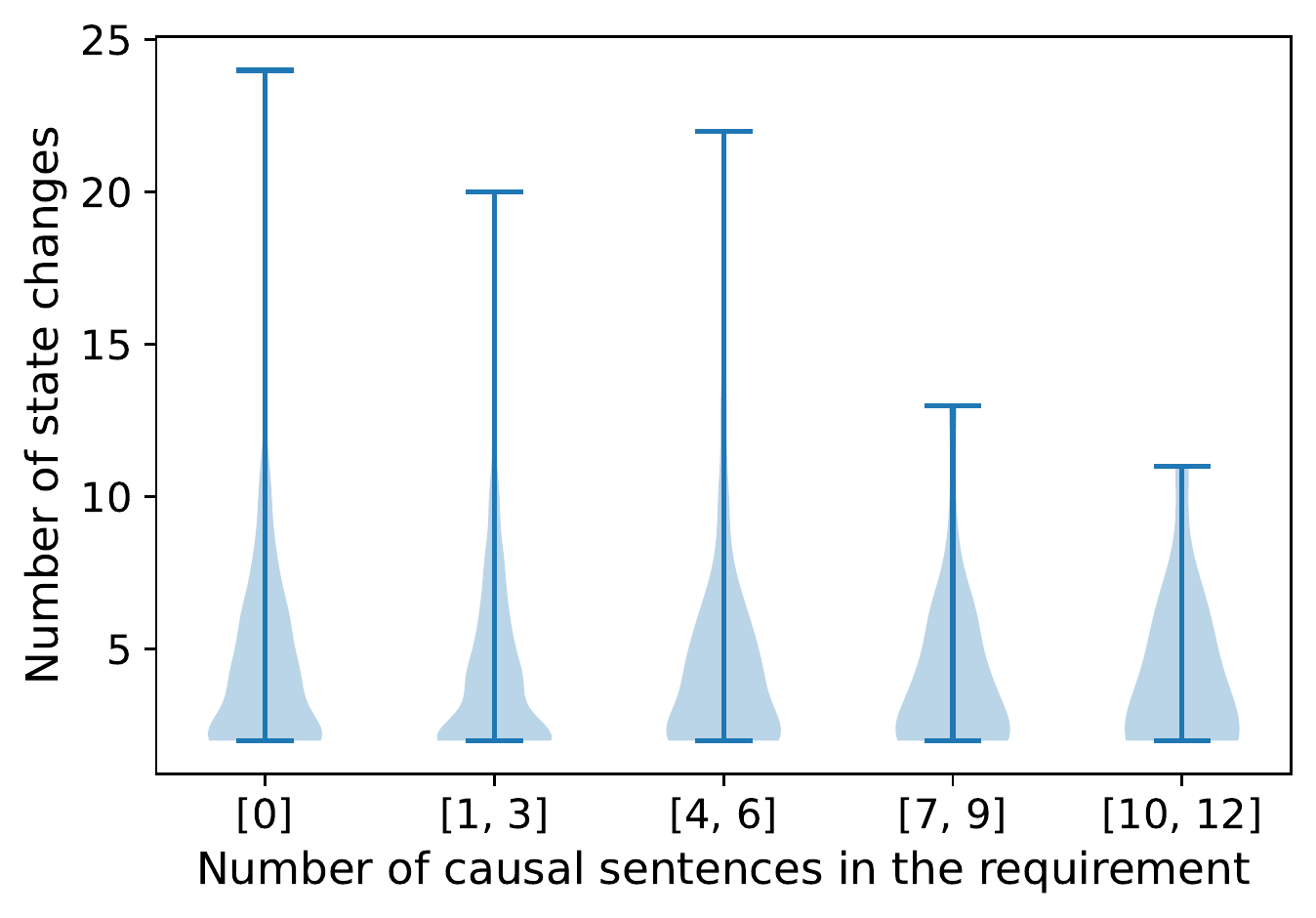}
        \caption{Batch granularity G2}
        \label{fig:cs2:h3g2}
    \end{subfigure}
    \caption{Distribution of volatility (H\textsubscript{3\textsubscript{0}})}
    \label{fig:cs2:h3}
\end{figure}

\begin{figure}
    \centering
    \begin{subfigure}{.32\textwidth}
        \includegraphics[width=\linewidth]{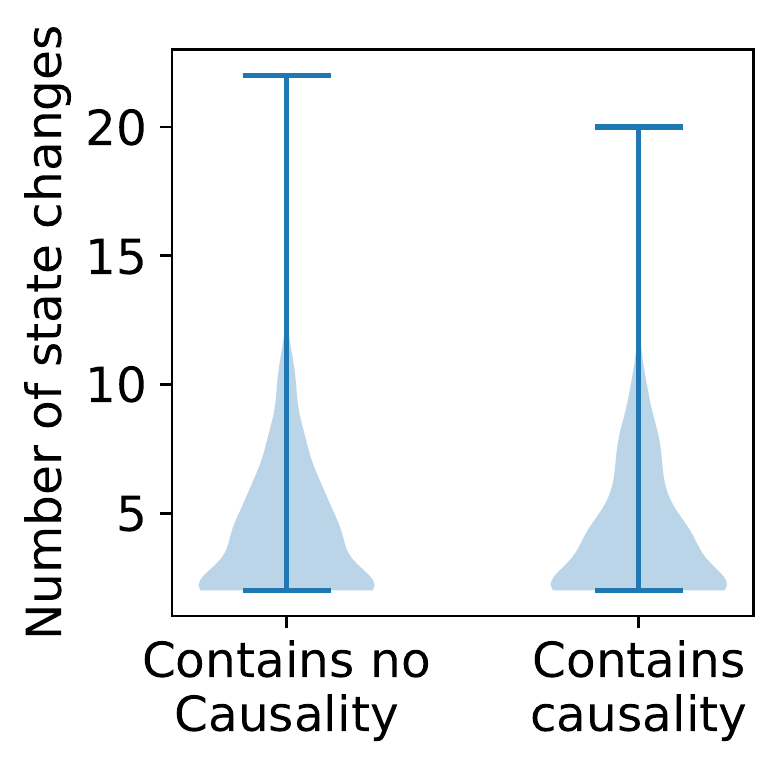}
        \caption{Small requirements}
        \label{fig:cs2:h3g3-small}
    \end{subfigure}
    \begin{subfigure}{.32\textwidth}
        \includegraphics[width=\linewidth]{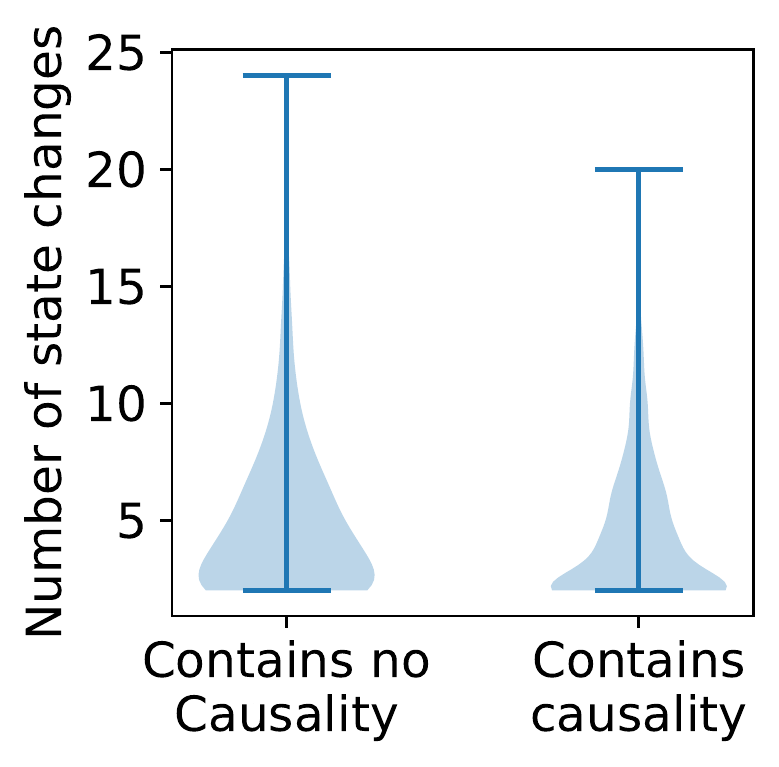}
        \caption{Medium requirements}
        \label{fig:cs2:h3g3-medium}
    \end{subfigure}
    \begin{subfigure}{.32\textwidth}
        \includegraphics[width=\linewidth]{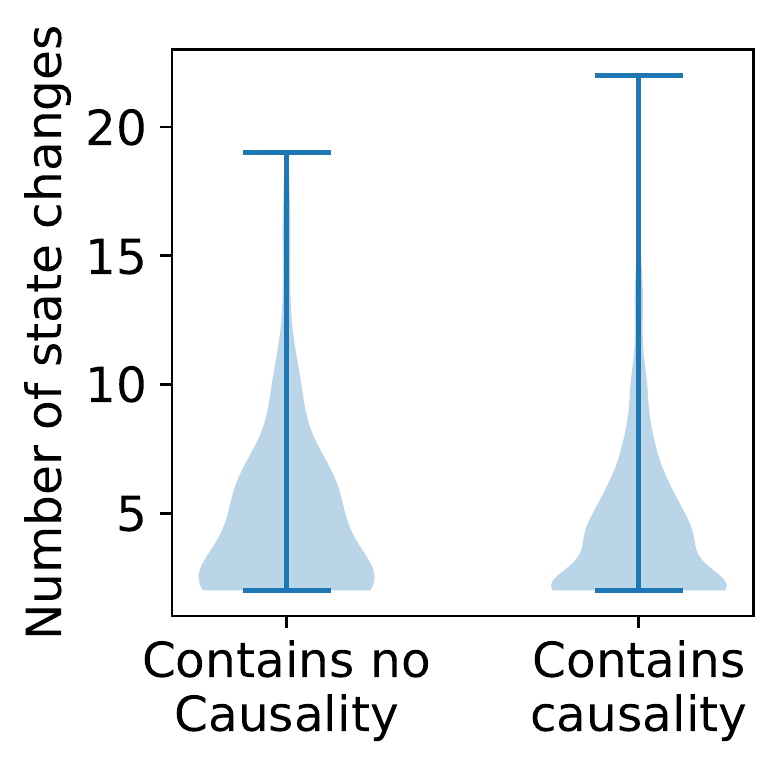}
        \caption{large requirements}
        \label{fig:cs2:h3g3-large}
    \end{subfigure}
    \caption{Distribution of volatility for binary granularity split by the size of requirements (H\textsubscript{3\textsubscript{0}})}
    \label{fig:cs2:h3g3}
\end{figure}

\paragraph{Correlation between causality and volatility}
Fig.~\ref{fig:cs2:h3g1} displays the distribution of the volatility metric in the two binary groups as violin plots. Overall, a slight correlation between the occurrence of causality and the volatility of a requirement can be observed, which is corroborated by the rejected test of independence with a p-value of $0.01$ and an effect size of 0.07. Investigating this effect at batch granularity G2 in Fig.~\ref{fig:cs2:h3g2}, however, reveals that this positive correlation is constrained to requirements with a low occurrence of causal sentences, where requirements with many causal sentences show a trade-off of higher average volatility despite a smaller overall range of volatility values. The Kruskal-Wallis test of independence does not allow to reject the null-hypothesis of similar distribution with a p-value of $0.086$. Investigating the correlation at granularity G3 as displayed in Fig.~\ref{fig:cs2:h3g3}, this trade-off is again visible and emphasizes the positive correlation between the occurrence of causality and the volatility of requirements for medium-sized requirements. This is confirmed by the independence tests, where the null hypothesis can be rejected for medium requirements with a p-value of $0.001$ and effect size of 0.18, but not for small or large requirements with p-values of $0.22$ and $0.13$ respectively.

\subsection{Implications}
\label{sec:effects:implications}

\paragraph{Impact of causality} 
The results of the second case study show an already existing positive correlation between the occurrence of causal relations and the features of requirements artifacts. These results motivate further, in-depth investigations corroborating the relationship between the occurrence of causality and features of these requirements, suggesting the feasibility of considering causality as an aspect of requirements quality. Both the direct insights and the consequent hypothesis for future research are discussed in more detail in the following paragraphs. 

\textbf{Answer to RQ 7:} The use of causality correlates slightly with smaller lead-times of requirements and therefore suggests considering causality as an impact factor when estimating the life-cycle of a requirement. A consequent hypothesis of this correlation is that the strict semantic structure of the relation causes an effect on the comprehensibility of a requirement, which makes it easier to translate into downstream artifacts like code or test cases. 

\textbf{Answer to RQ 8:} The use of causality does not correlate with the consolidated state of requirements. It is safe to assume that the occurrence of causality does not impact the life-cycle of requirements regarding its consolidated state statistically significant in comparison to other factors. 

\textbf{Answer to RQ 9:} The use of causality correlates slightly with smaller volatility of requirements. Comparably to RQ 7, the hypothesis that the semantic structure of the relation causes an effect on the understandability of a requirement, which in turn requires fewer decisions due to being less ambiguous, can be derived from the correlation analysis. We conclude that the relationship between the use of causality in NL requirements and lead-time as well as the volatility of requirements is worth for further, more thorough investigation: the slight correlation supports the feasibility of considering causality as an aspect of requirements quality.

\paragraph{Applicability of automatic causality detection}
The initial, exploratory investigation of this phenomenon demonstrates a possible use case of automatic causality detection as part of a quality metric. The small extent of the correlations and their low effect size according to the applied measures emphasize that the occurrence of causality is definitely not the only or most impactful, but certainly a considerable factor for the features of requirements. Considering the detection of causality with the approach presented in this research endeavor as a complement to other requirements quality frameworks such as for example requirements smells~\cite{femmer2017rapid} might benefit the reliability of these quality metrics by taking positive effects on requirements into account. Future studies need to investigate this claim in further detail.

\subsection{Threats to Validity}
\label{sec:effects:threats}

\paragraph{External validity}
The generalizability of results cannot be claimed based on the exploratory case study on one data set. Further data sets are necessary to be investigated and compared to compensate for context factors like the size and domain of the company, the utilized development process, techniques employed in the requirements engineering phase, as well as applied technologies. However, the analyzed data-set represents five years of product development (between 2010 and 2015) with 41 products and 36 software releases. Therefore, the heterogeneity of authors and editors of requirements is high.  
\paragraph{Internal validity}
To ensure that the correlation between the occurrence of causality and the lead-time of requirements is indeed causal and not confounded, further qualitative analysis beyond the data recorded in the respective data set must be performed. The impact of causality on comprehensibility in contrast to other factors of ambiguity has to be addressed in future studies of qualitative nature.
Apart from that, another possible threat to validity is that the analyzed data could contain incorrect information, caused for example by a lack of diligence when providing certain information for a requirement. Since this data set is based on real project data and has been fostered over the course of five years in an industrial setting the threat is considered low, but still worth mentioning. 

\section{Related Work}
\label{sec:related}

\subsection{Application of Causality Detection}

As indicated in Sect.~\ref{sec:terminology}, many disciplines have already dealt with the notion of causality and explored use cases for its application. One of the earliest applications of causality detection includes the utilization of causality for question answering. Girju et al.~\cite{girju2003automatic} propose an approach using lexico-syntactic patterns within one sentence or two adjacent sentences, where the patterns consist of two noun phrases (NP) connected with a causative verb (VP) in the following structure:
\begin{equation} <NP_1 verb NP_2> \end{equation}
The patterns were built by traversing WordNet concepts for noun phrases that are connected by a \textit{cause-to}-relationship, which is explicitly annotated in the WordNet corpus. Subsequently, from a large NL corpus, all verbs connecting these causally related noun phrase pairs were extracted as causation-verbs. Based on this information and further semantic features from WordNet, the lexico-syntactic patterns detecting causal relations were created. Chang et al.~\cite{chang2004causal} expand on this concept by taking into account conceptual pair probability and cue phrase probability as additional indicators for the classification of a causal relation. The focus on extracting causal relationships to automatically answer why-questions is expanded to the inter-sentential level by Pechsiri et al.~\cite{pechsiri2007mining}, who utilize the coexistence of causative and effective verbs as indicators for causal relationships. Other early approaches are rooted in the medical domain, where relationships between symptoms and diseases are commonly expressed in natural language sentences utilizing causality: Khoo et al.~\cite{khoo-etal-2000-extracting} extract causal knowledge from a medical database using graphical patterns. The roles and attributes of a causal situation are structured in a three-layer template, which constitutes the framework for manually elicited patterns. More recent approaches like the one proposed by Doan et al.~\cite{doan2019extracting} utilize POS tags and dependency parse trees to identify causal relations based on a manually generated set of patterns from a large data set of tweets. In the field of economics, causality detection has been applied to improve the reasoning about market-related relationships. Early approaches include Chan et al.~\cite{chan2005extracting} utilizing a hierarchy of manually generated semantic, sentence, and consequence and reason templates. Other approaches like proposed by Inui et al.~\cite{inui2005acquiring}, which also extract causal relations from newspapers, base their causality detection algorithm on the occurrence of cue phrases. The typology defined in the course of this research classifies causal relations with respect to their arguments' \textit{volitionality}, where the volitionality of an event distinguishes an action from a state of affairs. The resulting binary combinations of events of different volitionality constitute the four relationships \textit{cause}, \textit{effect}, \textit{precond}, and \textit{means}. Recent work by Xu et al.~\cite{xu2020review} acknowledges the lack of focus regarding labeling and extraction methods in the area of causality extraction and contributes by summarizing and evaluating existing causality data sets. Further applications of causality detection include extrapolating causal relations based on semantic relations between nouns~\cite{hashimoto2015generating}, effectively increasing the domain of reasoning based on causal relationships.

\subsection{Causality in Requirements Engineering}

To the best of our knowledge, we are the first to focus on causality from the perspective of Requirements Engineering. In our previously published papers, we motivated why the RE community should engage with causality~\cite{fischbach2020} and provide empirical evidence for the relevance of causality in requirement documents as well as further insights into its form and complexity~\cite{fischbach2021automatic}. The latter work is extended in the manuscript at hand with an additional, exploratory investigation of the implications of the use of causality in requirements artifacts.  Detecting causality in natural language has been investigated by several studies which usually belong to one of two categories according to Asghar et al.~\cite{Asghar16}: early approaches~\cite{khoo-etal-2000-extracting,wu05} use handcrafted, lexico-syntactic patterns to identify causal sentences. These approaches are highly dependent on the manually created patterns and show weak performance, inhibiting an effective application in practice as shown in our comparison of algorithms in Sect.~\ref{sec:detection}. Opposed to pattern-matching are feature-based classification methods: recent papers apply neural networks and exploit -- similarly to our approach -- the Transfer Learning capability of BERT~\cite{kyriakakis2019transfer}. However, we see a number of problems with these papers regarding the realization of our described RE use cases: First, neither the code nor a demonstration is published, making it difficult to reproduce the results and test the performance on data from the RE domain. Second, they train and evaluate their approaches on strongly unbalanced data sets with causal to non-causal ratios of 1:2 and 1:3, but only report the macro-Recall and macro-Precision values and not the metrics per class. Thus, it is not clear whether the classifier has a bias towards the majority class or not.

\section{Conclusions and Future Work}
\label{sec:conclusion}

The behavior of systems is often specified in terms of causal relations in natural language requirements. Efficiently extracting this causal information would allow for effective support of downstream activities that rely on such causal relations, such as the tool-supported derivation of test cases and further activities that need to reason about requirement dependencies~\cite{fischbach2020}. However, contemporary methods still fail to extract causality with reasonable performance~\cite{Asghar16}. Therefore, we have argued for the need for a novel method for causality extraction and closed this gap with the contributions in this manuscript.
We understand causality extraction as a two-step problem: First, we need to detect if requirements have causal properties. Second, we need to comprehend and extract their causal relations. At present, however, we lack knowledge about the form and complexity of causality in requirements, which is needed to develop suitable approaches for these two problems. In this manuscript, we reported on how we addressed this research gap by contributing: (C 1) an exploratory case study with 14,983 sentences from 53 requirements documents originating from 18 different domains. We found that causality is a widely used linguistic pattern to describe system functionalities and that it mainly occurs in explicit, marked form. (C 2) CiRA as an approach for the automatic detection of causality in requirements documents. This constitutes the first step towards causality extraction from NL requirements. We empirically evaluate our approach and achieve a macro-F\textsubscript{1} score of 82~\% on real-world data. (C 3) A demonstration of a possible use case of the automatic causality detection approach in a correlation analysis between the occurrence of causality and the life-cycle features of a requirement. (C 4) Finally, by following the open science norms and principles established in the empirical software engineering research community~\cite{mendez2019open}, we have further disclosed our entire source code, tool, and annotated data set within the limitations of existing non-disclosure agreements in order to actively support the research community working on same or similar problems and further facilitate independent replications.

Two further research directions are, in our opinion, worth being mentioned here: First, extending the first case study and analyzing the sentences from the requirements documents in a more granular way by categorizing them -- e.g., in functional and non-functional requirements -- would enrich our current insight into causality in requirements documents in general with further insights into causality in specific requirement categories. This includes investigating the particularities of specific domains, for example to explain the difference in cue phrase precision. Second, tackling the second of the two earlier mentioned sub-problems -- the actual extraction of causal relations from causal sentences -- will provide the necessary foundation to enable the various use cases. We are currently enhancing our previous approaches~\cite{fischbach20,frattini2020} with the insights gained from this study and cordially invite the RE community to join the endeavor. Building on the second case study presented in Sect.~\ref{sec:effects}, future studies may continue exploring the relationship between the occurrence of causality and features of requirements. Extending the automatic causality detection approach beyond the current intra-sentential limitations may for example enable to investigate the relationship between requirements' dependencies and features of requirements.

\begin{acknowledgements}
We would like to acknowledge that this work was supported, in parts, by the KKS foundation through the S.E.R.T. Research Profile project at Blekinge Institute of Technology. Further, we thank Yannick Debes and Michael Dorner for stimulating discussions and their valuable feedback on earlier versions of this work.
\end{acknowledgements}

% Authors must disclose all relationships or interests that 
% could have direct or potential influence or impart bias on 
% the work: 
%
% \section*{Conflict of interest}
%
% The authors declare that they have no conflict of interest.

%\bibliographystyle{spbasic}      % basic style, author-year citations
\bibliographystyle{spmpsci}      % mathematics and physical sciences
\bibliography{references}   % name your BibTeX data base
\clearpage

\begin{appendices}
\section{Detailed distribution of labels}

\begin{table}[!htbp]
\centering
 \rotatebox{90}{%
   \begin{minipage}{0.9\textheight}
\caption{Distribution of labels among all categories for all domains}
\label{tab:distribution:overview}
\resizebox{\textwidth}{!}{\begin{tabular}{l|rr|rr|rr|rr|rr|rr|rr|rrr|rrr|r}
    \toprule
    {} & \multicolumn{2}{c|}{Causal} & \multicolumn{2}{c|}{Explicit}  & \multicolumn{2}{c|}{Marked} & \multicolumn{2}{c|}{SingleCause} & \multicolumn{2}{c|}{SingleEffect}  & \multicolumn{2}{c|}{EventChain} & \multicolumn{2}{c|}{SingleSen} & \multicolumn{3}{c|}{Temporality} & \multicolumn{3}{c|}{Relationship} & Sentences\\
    {} & 0 & 1 & 0 & 1 & 0 & 1 & 0 & 1 & 0 & 1 & 0 & 1 & 0 & 1 & before & overlap & during & cause & enable & prevent & \\
    Domain          &             &             &               &               &             &             &                  &                  &                   &                   &                 &                 &                              &                              &                  &                  &                  &                    &                   &                   &           \\
    \midrule
    Aerospace       &        3846 &        1664 &           193 &          1471 &         296 &        1368 &              334 &             1330 &               380 &              1284 &            1534 &             130 &                         1550 &                          114 &              919 &              610 &              135 &               577 &               268 &                21 &       5510 \\
Agriculture     &         238 &          57 &             4 &            53 &           1 &          56 &               12 &               45 &                22 &                35 &              55 &               2 &                           56 &                            1 &               33 &               20 &                4 &                25 &                 5 &                 1 &        295 \\
Astronomy       &         133 &         106 &             5 &           101 &          27 &          79 &               21 &               85 &                27 &                79 &              94 &              12 &                          102 &                            4 &               34 &               67 &                5 &                40 &                13 &                 3 &        239 \\
Automotive      &         119 &          38 &             1 &            37 &           0 &          38 &               15 &               23 &                 9 &                29 &              35 &               3 &                           38 &                            0 &               24 &               13 &                1 &                18 &                 2 &                 1 &        157 \\
Banking         &         237 &         139 &             9 &           130 &          19 &         120 &               18 &              121 &                18 &               121 &             127 &              12 &                          133 &                            6 &               43 &               86 &               10 &                44 &                18 &                 0 &        376 \\
Data Analytics  &        2028 &         671 &            56 &           615 &         129 &         542 &              105 &              566 &               111 &               560 &             634 &              37 &                          629 &                           42 &              345 &              266 &               60 &               177 &                62 &                 6 &       2700 \\
Digital Library &         143 &          55 &             7 &            48 &           2 &          53 &                8 &               47 &                 9 &                46 &              54 &               1 &                           51 &                            4 &               23 &               27 &                5 &                12 &                10 &                 0 &        198 \\
Digital Society &          63 &          44 &             8 &            36 &          14 &          30 &                5 &               39 &                 3 &                41 &              41 &               3 &                           41 &                            3 &                0 &               39 &                5 &                12 &                 6 &                 0 &        107 \\
Energy          &          19 &           3 &             0 &             3 &           1 &           2 &                0 &                3 &                 0 &                 3 &               2 &               1 &                            3 &                            0 &                1 &                2 &                0 &                 2 &                 0 &                 0 &         22 \\
Health          &         784 &         378 &            26 &           352 &          56 &         322 &               58 &              320 &                82 &               296 &             349 &              29 &                          353 &                           25 &              167 &              169 &               42 &               105 &                50 &                 1 &       1164 \\
Infrastructure  &         498 &         278 &            26 &           252 &          38 &         240 &               58 &              220 &                48 &               230 &             262 &              16 &                          256 &                           22 &              140 &              118 &               20 &               100 &                12 &                 4 &        776 \\
Insurance       &          11 &           1 &             0 &             1 &           0 &           1 &                0 &                1 &                 0 &                 1 &               1 &               0 &                            1 &                            0 &                1 &                0 &                0 &                 1 &                 0 &                 0 &         12 \\
Military        &          43 &           1 &             0 &             1 &           0 &           1 &                0 &                1 &                 0 &                 1 &               1 &               0 &                            1 &                            0 &                0 &                1 &                0 &                 1 &                 0 &                 0 &         44 \\
Physics         &         140 &          91 &             5 &            86 &           5 &          86 &               26 &               65 &                21 &                70 &              83 &               8 &                           90 &                            1 &               34 &               47 &               10 &                37 &                10 &                 0 &        232 \\
Regulatory      &          25 &           2 &             0 &             2 &           0 &           2 &                0 &                2 &                 0 &                 2 &               2 &               0 &                            2 &                            0 &                1 &                1 &                0 &                 1 &                 0 &                 0 &         27 \\
Smart City      &        1168 &         346 &            48 &           298 &          27 &         319 &               83 &              263 &                79 &               267 &             317 &              29 &                          307 &                           39 &              131 &              172 &               43 &               107 &                50 &                 6 &       1514 \\
Sustainability  &         848 &         184 &            36 &           148 &          11 &         173 &               36 &              148 &                42 &               142 &             180 &               4 &                          161 &                           23 &               81 &               84 &               19 &                85 &                23 &                 2 &       1032 \\
Telecomm        &         421 &         157 &            22 &           135 &          23 &         134 &               26 &              131 &                27 &               130 &             154 &               3 &                          153 &                            4 &               67 &               79 &               11 &                53 &                23 &                 3 &        578 \\
\midrule
Sum             &       10764 &        4215 &           446 &          3769 &         649 &        3566 &              805 &             3410 &               878 &              3337 &            3925 &             290 &                         3927 &                          288 &             2044 &             1801 &              370 &              1397 &               552 &                48 &      14983 \\
    \bottomrule
    \end{tabular}}
\end{minipage}}
\end{table}
\end{appendices}

\end{document}